\documentclass[prb,aps,twocolumn,eqsecnum,amssymb,amsmath,showpacs]{revtex4}

\usepackage{epsfig}% Include figure files
\usepackage{graphicx}% Include figure files
\usepackage{dcolumn}% Align table columns on decimal point
\usepackage{bm}% bold math

\begin{document}

\title{       Orbital liquid in ferromagnetic manganites:\\
              The orbital Hubbard model for $e_g$ electrons }

\author {     Louis Felix Feiner }

\affiliation{ Institute for Theoretical Physics, Utrecht University,
              Leuvenlaan 4, NL-3584 CC Utrecht \\
              Philips Research Laboratories, Prof. Holstlaan 4,
              NL-5656 AA Eindhoven, The Netherlands }

\author {     Andrzej M. Ole\'{s} }

\affiliation{ Marian Smoluchowski Institute of Physics, Jagellonian
              University, Reymonta 4, PL-30059 Krak\'ow, Poland \\
              Max-Planck-Institut f\"ur Festk\"orperforschung,
              Heisenbergstrasse 1, D-70569 Stuttgart, Germany }

%\date{18 January, 2005}
%\date{\today}
\date{15 July, 2004}

\begin{abstract}
We have analyzed the symmetry properties and the ground state of an
orbital Hubbard model with two orbital flavors, describing a partly
filled spin-polarized $e_g$ band on a cubic lattice, as in ferromagnetic
manganites. We demonstrate that the off-diagonal hopping responsible for
transitions between $x^2-y^2$ and $3z^2-r^2$ orbitals, and the absence
of SU(2) invariance in orbital space, have important implications. One
finds that superexchange contributes in all orbital ordered states, the
Nagaoka theorem does not apply, and the kinetic energy is much enhanced
as compared with the spin case. Therefore, orbital ordered states are
harder to stabilize in the Hartree-Fock approximation (HFA), and the
onset of a uniform ferro-orbital polarization and antiferro-orbital
instability are similar to each other, unlike in spin case. Next we
formulate a cubic (gauge) invariant slave boson approach using the
orbitals with complex coefficients. In the mean-field approximation it
leads to the renormalization of the kinetic energy, and provides a
reliable estimate for the ground state energy of the disordered state.
Using this approach one finds that the HFA fails qualitatively in the
regime of large Coulomb repulsion $U\to\infty$ --- the orbital order
is unstable, and instead a strongly correlated {\it orbital liquid\/}
with disordered orbitals is realized at any electron filling.\\
{\it [Published in: Phys. Rev. B {\bf 71}, 144422 (2005).]}
\end{abstract}


\maketitle

%%%%%%%%%%%%%%%%%%%%%%%%%%%%%%%%%%%%%%%%%%%%%%%%%%%%%%%%%%%%%%%%%%%%%%%%
%%
%%                            Introduction
%%
%%%%%%%%%%%%%%%%%%%%%%%%%%%%%%%%%%%%%%%%%%%%%%%%%%%%%%%%%%%%%%%%%%%%%%%%
\section{Introduction}
\label{sec:intro}

In recent years there has been renewed interest in orbital degrees of
freedom in Mott insulators.\cite{Tok00} Typically, Mott insulators are
stoichiometric, i.e., {\it undoped\/}, oxides (or sulfides) in which
the strong on-site interorbital Coulomb repulsion $U$ on the
transition-metal ions dominates over the kinetic energy driven by the
electron hopping $t$, and eliminates charge fluctuations. At energies
well below $U$ one is then left with effective low-energy interactions
$\propto t^2/U$ of superexchange (SE) type.
In many cases these are purely magnetic interactions between the
spins on the metal ions, leading to the familiar spin models, such as
the Heisenberg model. However, when the electrons occupy partly-filled
degenerate $e_g$ or $t_{2g}$ orbitals, such as in the perovskites
KCuF$_3$, LaMnO$_3$, LaTiO$_3$, and LaVO$_3$, the orbital degrees of
freedom become equally important as the spin ones, and it is therefore
necessary to treat both of them on equal footing. In such cases the SE
is described by so-called spin-orbital models,\cite{Kug82,Ole03} and
the SE interactions are typically strongly frustrated, even on a cubic
lattice.\cite{Fei97} In spin-orbital models the quantum effects are
particularly strong --- the quantum fluctuations are enhanced, and
might even destabilize the long-range magnetic order, leading to a
{\it spin liquid\/} state, possibly realized in LiNiO$_2$.\cite{Fei97}
The opposite situation, that an (isotropic or anisotropic)
{\it orbital liquid\/} (OL) is stabilized and coexists with long-range
spin order, was pointed out recently for $t_{2g}$ Mott-Hubbard systems.
\cite{Kha00} By contrast, in undoped $e_g$ systems, such as KCuF$_3$
(Ref. \onlinecite{Kug82}) and LaMnO$_3$ (Ref. \onlinecite{Fei99}), the
quantum phenomena are partly quenched and the SE favors alternating
orbital (AO) order which coexists with antiferromagnetic (AF) spin order.

An issue of considerable interest is how such systems, characterized by
the presence of orbital degrees of freedom, behave under doping,
and in particular how this compares with the more familiar behavior of
doped spin systems. In this paper we address this issue by studying a
generic model of correlated $e_g$ electrons with two orbital flavors,
described by a pseudospin $T=1/2$ in the orbital Hilbert space,
and consider its relation to the standard (spin) Hubbard model for
electrons with spin $S=1/2$.
So we introduce the {\it $e_g$-orbital Hubbard model\/} and investigate:
 (i) in what respect long-range order in such an {\it orbital system\/}
     is different from that in the analogous {\it spin system\/}, and
(ii) whether the orbitals may order when $U$ is large, or rather
     form a {\it disordered\/} OL.
These questions are of fundamental nature and our main aim in addressing
them is to uncover and elucidate the {\it physical mechanisms\/} which
operate in the $e_g$ band and are typical for orbital degeneracy, in
particular by contrasting them with those known to operate in spin
systems.

The present problem is closely related to the physical properties of
the colossal magnetoresistance (CMR) manganites,\cite{Ima98} where the
well-known mechanism of double exchange introduced by Zener\cite{Zen51}
is responsible for the metallic ferromagnetic (FM) phase at finite
doping, in which the spins of the $e_g$ electrons are fully polarized.
The model that we will investigate here covers only the case of the FM
phase, thus neglecting the competition of the double-exchange mechanism
with the spin AF SE, and the resulting dependence of the
hopping amplitude on the actual spin states at two neighboring sites.
\cite{Fes01} Even when one limits oneself to the FM phase, a sequence of
orbital-ordered phases may be expected,\cite{Mae98} and each of them
would break cubic symmetry, contrary to what is observed in the magnetic
properties of the metallic FM phase. The analysis of the present paper,
making an extensive use of the auxiliary particle method in the strongly
correlated regime,\cite{Bar76} provides the basis for a proper treatment
of this problem,\cite{Ole02} which enables one to understand the
persistence of cubic symmetry and more in particular why the magnon
stiffness constant increases with hole doping.\cite{End97}

This paper is organized as follows. In Section \ref{sec:ohm} we
introduce the orbital Hubbard model for spin-polarized $e_g$ electrons
at orbital degeneracy, and discuss its symmetry properties. We show
that the cubic symmetry of the hopping may be better appreciated when
a particular basis consisting of two orbitals with complex coefficients
is used. Next we analyze in Sec. \ref{sec:oos} the possible orbital
ordered phases at $U=\infty$ and compare their densities of states and
total energies derived within the slave fermion formalism. Such phases
follow from the instabilities towards orbital-ordered states obtained
within the Hartree-Fock (HF) approximation (Sec. \ref{sec:hf}), and we
show that such instabilities and the properties of the ordered phases at
finite $U$, related to the SE, are here quite different from those
known from the spin Hubbard model.
In Sec. \ref{sec:oliq} we introduce the cubic invariant slave boson
approach and use the mean-field approximation to analyze the disordered
orbital liquid state. Within a generalization of the Kotliar-Ruckenstein
\cite{Kot86} (KR) approach to the present orbital problem, we give
reasons why the orbital ordered states are unstable against the
OL disordered state when one goes beyond the HF approximation.
The paper is concluded in Sec. \ref{sec:summa} by pointing out the
implications of our results for the physical properties of the
CMR manganites.

%%%%%%%%%%%%%%%%%%%%%%%%%%%%%%%%%%%%%%%%%%%%%%%%%%%%%%%%%%%%%%%%%%%%%%%%
%%
%%                        Orbital Hubbard Model
%%
%%%%%%%%%%%%%%%%%%%%%%%%%%%%%%%%%%%%%%%%%%%%%%%%%%%%%%%%%%%%%%%%%%%%%%%%
\section{Orbital Hubbard Model}
\label{sec:ohm}

\subsection{The Hamiltonian and its symmetry properties}

We consider spinless $e_g$ electrons on a cubic lattice with kinetic
energy
\begin{equation}
H_t=-t \sum_{\alpha} \sum_{\langle ij\rangle\parallel\alpha}
    c_{i\zeta_{\alpha}}^{\dagger}c_{j\zeta_{\alpha}}^{},
\label{H_zeta}
\end{equation}
where hopping with amplitude $-t$ between sites $i$ and $j$ occurs only
for a pair of directional orbitals $|\zeta_{\alpha}\rangle$ oriented
along the bond $\langle ij\rangle$ direction, i.e.,
$|\zeta_{\alpha}\rangle\propto 3x^2-r^2$, $3y^2-r^2$, and $3z^2-r^2$,
when the bond $\langle ij\rangle$ is along the cubic axis $\alpha=a$,
$b$, and $c$, respectively. We will similarly denote by
$|\xi_{\alpha}\rangle$ the orbital which is orthogonal to
$|\zeta_{\alpha}\rangle$ and is oriented perpendicular to the bond
$\langle ij\rangle$, i.e., $|\xi_{\alpha}\rangle\propto y^2-z^2$,
$z^2-x^2$, and $x^2-y^2$, for a bond $\langle ij\rangle$ along
the axis $\alpha=a$, $b$, and $c$, respectively.
While such a choice of basis, that depends on the bond direction under
consideration, is convenient for writing down the kinetic energy, one
cannot avoid to choose a particular orthogonal basis for the two
orbital flavors as soon as one wants to introduce a Hubbard term to
describe the local electron interactions. The usual choice is to take
\begin{equation}
\label{realorbs}\textstyle{
|z\rangle\equiv \frac{1}{\sqrt{6}}(3z^2-r^2),
\hspace{0.7cm}
|x\rangle\equiv \frac{1}{\sqrt{2}}(x^2-y^2),}
\end{equation}
called {\it real orbitals\/}.
However, because this basis is the natural one only for the bonds
parallel to the $c$ axis but not for those in the $(a,b)$ plane, the
kinetic energy then takes the form \cite{Tak98,Bri99}
\begin{eqnarray}
H_t\! &=&\! -\frac{1}{4}t\!\sum_{\langle ij\rangle\parallel a,b}
  \big[3c_{ix}^{\dagger}c_{jx}^{}+c_{iz}^{\dagger}c_{jz}^{}
  \mp\sqrt{3} (c_{ix}^{\dagger}c_{jz}^{}+c_{iz}^{\dagger}c_{jx}^{})\big]
                                                  \nonumber  \\
    & &\! -t\sum_{\langle ij\rangle\parallel c}
    c_{i z}^{\dagger}c_{j z}^{},
\label{H_real}
\end{eqnarray}
and although this expression is of course cubic invariant, the
representation (3) of the hopping does not exhibit this symmetry but
takes a very different appearance depending on the bond direction.

We thus prefer to use instead the basis of {\it complex orbitals\/} at
each site\cite{noteJTconv}
\begin{equation}\textstyle{
|+\rangle=\frac{1}{\sqrt{2}}\big(|z\rangle - i |x\rangle\big),
\hspace{0.5cm}
|-\rangle=\frac{1}{\sqrt{2}}\big(|z\rangle + i |x\rangle\big),}
\label{complex}
\end{equation}
corresponding to ``up'' and``down'' pseudospin flavors, with the local
pseudospin operators defined as
\begin{eqnarray}
T_i^+&=&c_{i +}^{\dagger} c_{i -}^{},    \hspace{1cm}
T_i^- = c_{i -}^{\dagger} c_{i +}^{},    \nonumber \\
T_i^z&=&\textstyle{\frac{1}{2}}
  (c_{i +}^{\dagger} c_{i +}^{} - c_{i -}^{\dagger} c_{i -}^{} )
= \textstyle{\frac{1}{2}}(n_{i +} - n_{i -}).
\label{pseudospin}
\end{eqnarray}
For later reference it is convenient to introduce also electron
creation operators $c_i^{\dagger}(\psi_i,\theta_i)$ which create $e_g$
electrons in orbital coherent states, defined as
\begin{eqnarray}
|\Omega_i\rangle =
 e^{-i\theta_i/2}\cos\Big(\frac{\psi_i}{2}\Big)|i+\rangle
+e^{+i\theta_i/2}\sin\Big(\frac{\psi_i}{2}\Big)|i-\rangle,
\label{cohorb}
\end{eqnarray}
in analogy with the well-known spin coherent states.\cite{Kla79}
The expectation value of the local pseudospin operator in the coherent
orbital (\ref{cohorb}) behaves like a classical vector,\cite{notecoh}
\begin{equation}
\langle \Omega_i| {\bm T}_i |\Omega_i\rangle =
\textstyle{\frac{1}{2}} (\sin\psi_i\cos\theta_i,
                         \sin\psi_i\sin\theta_i, \cos\psi_i ),
\label{vector}
\end{equation}
traversing a sphere, with the ``equatorial plane'' ($\psi_i=\pi/2$)
corresponding to real orbitals
$|\Omega_i (\pi/2,\theta_i)\rangle \equiv |i\theta_i\rangle =
\cos(\theta_i/2) |iz\rangle - \sin(\theta_i/2) |ix\rangle$, and
the ``poles'' ($\psi_i=0$ and $\psi_i=\pi$) to the complex orbitals
$|i+\rangle$ and $|i-\rangle$. The three directional orbitals
$|i\zeta_{\alpha}\rangle$ at site $i$, associated with the three cubic
axes ($\alpha=a$, $b$, $c$), are the real orbitals with $\theta_i$
being equal to $\vartheta_a=-4\pi/3$, $\vartheta_b=+4\pi/3$, and
$\vartheta_c=0$, respectively, i.e.
\begin{eqnarray}
|i \zeta_{\alpha}\rangle &=&\textstyle{\frac{1}{\sqrt{2}}}
    [e^{-i\vartheta_{\alpha}/2}|i+\rangle
    +e^{+i\vartheta_{\alpha}/2}|i-\rangle]    \nonumber \\
  &=& \cos(\vartheta_{\alpha}/2)|iz\rangle
     -\sin(\vartheta_{\alpha}/2)|ix\rangle,
\label{zetaorb}
\end{eqnarray}
and thus correspond to the pseudospin lying in the equatorial plane and
pointing in one of the three equilateral ``cubic'' directions defined by
the angles $\vartheta_{\alpha}$.

In the complex-orbital representation (\ref{complex}) the {\it orbital
Hubbard model\/} for $e_g$ electrons takes the form
\begin{eqnarray}
\cal{H}&=& -\frac{1}{2} t \sum_{\alpha}
\sum_{\langle ij\rangle\parallel\alpha}
  \Big[\Big(c_{i+}^{\dagger}c_{j+}^{}+c_{i-}^{\dagger}c_{j-}^{}\Big)
                                \nonumber \\
  & & \hskip 1.7cm
  + \gamma \Big(e^{-i\chi_{\alpha}}c_{i+}^{\dagger}c_{j-}^{}
    +e^{+i\chi_{\alpha}}c_{i-}^{\dagger}c_{j+}^{}\Big)\Big]
                                \nonumber \\
  & & + U \sum_in_{i+}^{}n_{i-}^{},
\label{H_c}
\end{eqnarray}
with $\chi_a=+2\pi/3$, $\chi_b=-2\pi/3$, and $\chi_c=0$, and where the
newly introduced parameter $\gamma$, explained below, takes the value
$\gamma=1$. The appearance of the phase factors $e^{\pm i\chi_{\alpha}}$
is characteristic of the orbital problem --- they occur because the
orbitals have an actual shape in real space so that each hopping process
depends on the bond direction. The form of the interorbital Coulomb
interaction $\propto U$ is invariant under any local basis
transformation to a pair of orthogonal orbitals; it gives an energy $U$
either when two real orbitals are simultaneously occupied,
$U\sum_in_{ix}n_{iz}$, or when two complex orbitals are occupied, as in
Eq. (\ref{H_c}).

The representation (\ref{H_c}) has several advantages:
  (i) It displays manifestly the cubic symmetry, since the
transformation $\chi_{\alpha} \rightarrow \chi_{\alpha} + 2\pi/3$
(which amounts in Eq. (\ref{H_c}) to the cyclic permutation
$a\rightarrow b\rightarrow c\rightarrow a$ of the cubic axes) in
conjunction with the corresponding phase shift of the electron operators
$c_{i\pm}^{\dagger}\rightarrow c_{i\pm}^{\dagger} e^{\pm 2i\pi/3}$
(which permutes the $|\zeta_{\alpha}\rangle$-orbitals according to
$3x^2-r^2\rightarrow 3y^2-r^2\rightarrow 3z^2-r^2\rightarrow 3x^2-r^2$)
leaves the Hamiltonian (\ref{H_c}) invariant.
 (ii) It exhibits clearly the difference between the spin case and the
orbital case. In the orbital case there is both pseudospin-conserving
hopping [the first line in Eq.~(\ref{H_c})] and
{\it non-pseudospin-conserving hopping\/} [the second line in Eq.
(\ref{H_c})], whereas in the corresponding spin case, i.e., in the
standard Hubbard model, there is of course only spin-conserving hopping
and the second term is absent. Thus the present complex-orbital
representation allows us to introduce the parameter $\gamma$ by which
one can turn the $e_g$-band orbital Hubbard model ($\gamma=1$) into
what is formally a spin Hubbard model with the same hopping amplitudes
($\gamma=0$), interpreting ``$+$'' and ``$-$'' as ``spin up'' and
``spin down''. This device makes it very easy to recognize the
differences in physical behavior between the orbital case and the spin
case: the parameter $\gamma$ will of course show up in all analytical
expressions below, and one can compare at a glance the result for the
orbital case ($\gamma=1$) with that for the spin case ($\gamma=0$). At
the present stage one can already observe from Eq.~(\ref{H_c}) that
there is {\it more\/} kinetic energy available per electron in the
orbital case, because additional hopping channels are present. We will
see below that this has important consequences for the relative
stability of various states. (iii) Finally, it shows explicitly that
rotational SU(2) symmetry for the pseudospins is absent,\cite{Kug82}
which in the complex-orbital representation is immediately obvious
from the presence of the non-pseudospin-conserving hopping term
$\propto\gamma$ in (\ref{H_c}). Thus the components of the total
pseudospin operator ${\bm {\mathcal T}}=\sum_i {\bm T}_i$, are conserved
only at $\gamma=0$ (i.e., $[{\bm {\mathcal T}},{\cal H}]=0$), while the
terms $\propto\gamma$ in ${\cal H}$ commute instead with the staggered
pseudospin operator ${\cal T}^z_{\bf Q}=
\sum_i\exp(i{\bf Q \cdot R}_i)T_i^z$, where ${\bf Q}=(\pi,\pi,\pi)$.

\subsection{ New features compared with the spin case }

%%%%%%%%%%%%%%%%%%%%%%%%%%%%%%%%%%%%%%%%%%%%%%%%%%%%%%%%%%%%%%%%%%%%%%%%
%%
%%                             figure 1
%%
%%%%%%%%%%%%%%%%%%%%%%%%%%%%%%%%%%%%%%%%%%%%%%%%%%%%%%%%%%%%%%%%%%%%%%%%
\begin{figure}[t!]
\includegraphics[width=7.7cm]{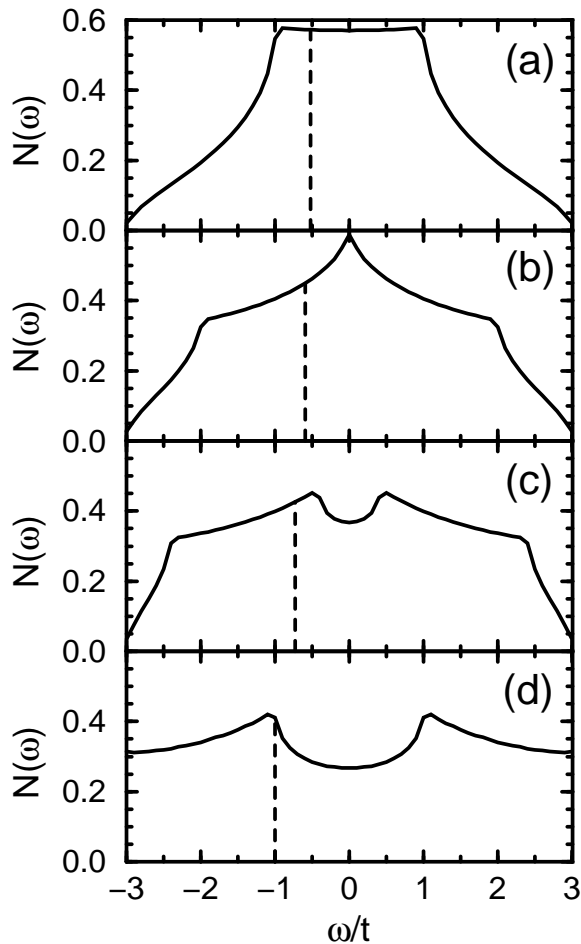}
\caption{
Evolution of the density of states $N(\omega)$ (in units of $t=1$)
obtained for the tight-binding model (\protect\ref{H_c}) at $U=0$ with
increasing off-diagonal hopping $\gamma t$:
(a) $\gamma=0$, (b) $\gamma=0.5$, (c) $\gamma=1/\sqrt{2}$,
(d) $\gamma=1$; $\gamma=0$ and $\gamma=1$ corresponds to the spin and
to the orbital Hubbard model, respectively.
Dashed lines show, for the same electron filling $n=0.7$ in all cases,
the Fermi energy, which decreases with increasing $\gamma$ for $n<1$.
}
\label{fig:dosg}
\end{figure}

It is instructive to follow the changes of the electronic structure of
the uncorrelated band [i.e. with $U=0$ in Eq.~(\ref{H_c})] with
increasing $\gamma$ ($0\le\gamma\le 1$). When the hopping is only
diagonal between
pairs of $|+\rangle$ and $|-\rangle$ states at $\gamma=0$, the
pseudospin bands are degenerate, and the density of states has the
familiar shape obtained for a simple cubic lattice, with bandwidth
$6t$, corresponding to the hopping elements of $\frac{1}{2}t$ [Fig.
\ref{fig:dosg}(a)]. Increasing $\gamma$ removes the degeneracy of the
electron bands, and gives increasing spectral weight near the band edges
without modifying the bandwidth. For genuine $e_g$ electrons (i.e. at
$\gamma=1$) the density of states does not start from zero at
$\omega=\pm 3t$, as usually for three-dimensional (3D) lattices, but is
finite there and has a value close to its average over the entire band
[Fig. \ref{fig:dosg}(d)]. Not only is the spectral weight transferred to
lower energies, but even the Fermi energy at fixed electron density
$n<1$ decreases with increasing $\gamma$, as shown on the example of
$n=0.7$ in Fig. \ref{fig:dosg}. Therefore, for a given electron density,
at $U=0$ the kinetic energy of $e_g$ electrons (i.e. at $\gamma=1$) is
{\it lower\/} than in the corresponding spin case (at $\gamma=0$).

Finally, some remarks on the physical interpretation of the orbital
Hubbard model are in place here. The first of them concerns electron
spin. As said, the electrons in the model are spinless [{\it cf.\/}
Eq.~(\ref{H_zeta})], which at first sight may seem unphysical. However,
such a model is entirely appropriate for real, i.e. spincarrying $e_g$
electrons in a FM state, where the spins are fully polarized. This
situation can be realized in a strong magnetic field, or, as in
manganites, when the double exchange polarizes the $t_{2g}$ core spins
which in turn polarize the $e_g$ band by strong Hund's rule coupling.
Then the spin degrees of freedom are completely frozen out and only the
orbital degrees of freedom remain and can contribute to the kinetic
energy. Actually, Eq.~(\ref{H_zeta}) [but with the additional constraint
of no double occupancy] is precisely the expression for the kinetic
energy of the $e_g$ band in the metallic ferromagnetic phase of the
doped manganites La$_{1-x}$A$_x$MnO$_3$ (with A = Sr, Ca,..., and
$x\sim 0.3$) when these are described by an extended (orbital-degenerate
and large spin) $t$-$J$ model.\cite{Ole02} So the $U\rightarrow \infty$
limit of the present orbital Hubbard model (\ref{H_c}) is directly
relevant for the physics of the manganites, and for this reason we will
pay extra attention to this limit.

The second remark concerns the parameter $\gamma$.
As one can readily verify, the kinetic terms in Eq.~(\ref{H_c}) with
arbitrary $\gamma$ are equivalent to the kinetic energy Hamiltonian
\begin{equation}
\hat{H}_t=-\frac{1}{2}t \sum_{\alpha}
    \sum_{\langle ij\rangle\parallel\alpha}
    \Bigl[ (1+\gamma) \: c_{i\zeta}^{\dagger}c_{j\zeta}^{}
         + (1-\gamma) \: c_{i\xi}^{\dagger}c_{j\xi}^{} \Bigr],
\label{H_zetaxi}
\end{equation}
which reduces to $H_t$ [Eq.~(\ref{H_zeta})] for $\gamma=1$.
So, although we have introduced the parameter $\gamma$
purely as a formal device, it actually describes the relative strength
of hopping between the $|\xi_{\alpha}\rangle$ orbitals perpendicular to
a bond, and one sees that $\gamma=1$ corresponds to ``$\zeta$-hopping
only'', $\gamma=0$ to ``$\zeta$-hopping and $\xi$-hopping equally
strong'' (equivalent to the spin case as discussed above), and
$\gamma=-1$ to ``$\xi$-hopping only''. Although such $\xi$-hopping
occurs, for instance, in transition metals as a $(dd\delta)$ element,
and is symmetry-allowed in the perovskites, it {\it cannot\/} occur by
the familiar mechanism of two-step hopping (neither $\sigma$-type nor
$\pi$-type) via a $2p$ orbital on the oxygen ion in between two
transition metal ions. It is therefore generally accepted that in
physically relevant cases this hopping process is smaller by at least
two orders of magnitude than that between $|\zeta_{\alpha}\rangle$
orbitals, and thus, to our knowledge, all work on the manganites has
actually been done assuming pure $\zeta$-hopping, i.e. $\gamma=1$.
Nevertheless, we will occasionally let $\gamma$ vary between 0 and 1,
not with the intention to suggest that a significant strength of
$\xi$-hopping is actually physically relevant, but rather with the
purpose of demonstrating how the non-pseudospin-conserving hopping
affects the physical properties of strongly correlated electrons in
a partly filled band.

The third remark concerns the difference between real and complex
orbitals. It is noteworthy that, unlike in the spin case, already for
an individual site there is no spherical symmetry in pseudospin space
even at the classical level: the directions available to the
pseudospinvector are {\it not\/} all physically equivalent. In
particular, the real orbitals are spatially anisotropic and have a
nonzero diagonal electric quadrupole moment (EQM), $\langle T^x_i
\rangle^2 +\langle T^y_i \rangle^2 \neq 0$, whereas the complex
orbitals have a cubic shape, with only $\langle T^z_i\rangle\neq 0$.
This difference is of course the origin for the hopping Hamiltonian
not having SU(2) symmetry. Moreover, as pointed out by Van den Brink
and Khomskii,\cite{Bri01} in a real compound like a perovskite the EQM
couples directly to the lattice, and occupancy of a real orbital would
induce a local Jahn-Teller (JT) distortion whereas occupancy of
a complex orbital would not.\cite{note:JT}

%%%%%%%%%%%%%%%%%%%%%%%%%%%%%%%%%%%%%%%%%%%%%%%%%%%%%%%%%%%%%%%%%%%%%%%%
%%
%%                         Orbital Ordered States
%%
%%%%%%%%%%%%%%%%%%%%%%%%%%%%%%%%%%%%%%%%%%%%%%%%%%%%%%%%%%%%%%%%%%%%%%%%
\section{Orbital Ordered States}
\label{sec:oos}

\subsection{Uniform and alternating orbital order}
\label{sec:core}

Because the electrons interact by the local Coulomb interaction $U$,
they are prone to instabilities towards {\it orbital order\/}, similar
to the magnetic instabilities towards spin order in the spin case,
\cite{Faz99} to which we will compare them. At half-filling ($n=1$) the
simplest possibility to reduce the interaction energy $\propto U$ would
be to polarize the system completely into {\it ferro orbital\/} (FO)
states,
\begin{equation}
|\Phi_{\rm FO}\rangle=\prod_i c_i^{\dagger}(\psi,\theta)|0\rangle,
\label{fo}
\end{equation}
with the pseudospin pointing in the same direction at all sites. As in
the spin case, another possibility is {\it alternating orbital\/} (AO)
order,
\begin{equation}
|\Phi_{\rm AO}\rangle =
   \prod_{i\in A} c_{i}^{\dagger}(\psi_A,\theta_A)
   \prod_{j\in B} c_{j}^{\dagger}(\psi_B,\theta_B)|0\rangle,
\label{afo}
\end{equation}
with orbitals alternating between two sublattices $A$ and $B$ which
cover a cubic lattice. Depending on whether orbitals alternate in every
direction, or whether there are lines or planes of ferro orbital order,
these states are classified as $G$-type (for spin called N\'eel states),
$C$-type, or $A$-type AO states. Doubly occupied sites are explicitly
avoided in all these states. If the band is partly
filled ($n<1$), these ordered states must of course be modified to
involve a coherent mixture of orbital-polarized occupied sites and
empty sites. Such fully polarized states are appropriate only in the
$U\to\infty$ limit, where double occupancy is fully suppressed by the
Hubbard term and only the kinetic energy,
$E_{\rm kin}=\langle H_t\rangle$, remains relevant.

In contrast to the spin case, where because of the SU(2) symmetry both
the FM spin state and the AF spin state are unique, in the present
orbital case without SU(2) symmetry there is already a plethora of
physically different ordered states even if one does not go beyond two
sublattices. In particular, as shown by Takahashi and Shiba,\cite{Shi00}
Maezono and Nagaosa,\cite{Mae00} Shen {\it et al.\/},\cite{She00}
and particularly stressed by Van den Brink and Khomskii,\cite{Bri01} it
makes a big difference whether one builds an ordered state completely
from complex orbitals (and empty sites) [i.e., $\psi$, $\psi_A$,
$\psi_B$ $=0,\pi$], leading to what we shall call {\it complex states\/},
or whether one uses exclusively real orbitals [i.e., $\psi$, $\psi_A$,
$\psi_B$ $=\pi/2$], thus constructing {\it real states\/}. This can be
conveniently demonstrated explicitly by formalizing the description of
the $U\to\infty$ limit by means of the slave fermion formalism, which
permits treatment of the general case (i.e., arbitrary $\psi$'s and
$\theta$'s). We present such states here in some detail, since the
$U\to\infty$ limit will serve as a reference in the later discussion.

So we introduce orbital bosons $b_{i \eta}^{\dagger}$ (with $\eta=+,-$)
to represent the occupied $e_g$ orbitals, $|\pm\rangle_i=
c_{i\pm}^{\dagger}|0\rangle\equiv b_{i\pm}^{\dagger}|{\rm vac}\rangle$,
and positively charged slave fermions $f_i^{\dagger}$ to represent the
empty sites, $|0\rangle_i \equiv f_i^{\dagger}|{\rm vac}\rangle$.
Thus the original electron operators are replaced according to
$c_{i \pm}^{\dagger}=b_{i \pm}^{\dagger} f_i^{}$,
and the Hamiltonian takes the form
\begin{eqnarray}
{\cal H}_{U=\infty} &=&
+\frac{1}{2} t \sum_{\alpha} \sum_{\langle ij\rangle\parallel\alpha}
  f_{i}^{\dagger}f_{j}^{}
  \Big[\Big(b_{i+}^{}b_{j+}^{\dagger}+b_{i-}^{}b_{j-}^{\dagger}\Big)
                                \nonumber \\
  &+ &\gamma \Big(e^{+i\chi_{\alpha}}b_{i+}^{}b_{j-}^{\dagger}
    +e^{-i\chi_{\alpha}}b_{i-}^{}b_{j+}^{\dagger}\Big)\Big],
\label{hsf}
\end{eqnarray}
with the local constraint
\begin{equation}
b_{i+}^{\dagger} b_{i+}^{} + b_{i-}^{\dagger} b_{i-}^{} +
f_i^{\dagger} f_i^{}=1,
\label{mixconst}
\end{equation}
implementing the condition of no double occupancy. Orbital order is
then imposed by treating the bosons in mean field approximation, i.e.
by making the replacements [compare Eq. (\ref{cohorb})],\cite{notecon}
\begin{eqnarray}
b_{i+} &\rightarrow& \cos(\psi_i/2)e^{-i\theta_i/2},    \nonumber \\
b_{i-} &\rightarrow& \sin(\psi_i/2)e^{+i\theta_i/2},
\label{bpm}
\end{eqnarray}
upon which the local pseudospin operators ${\bm T_i}$ are given by
Eq.~(\ref{vector}). We are then left with a Hamiltonian describing
fermionic holes moving in a background of fixed orbitals.

In the case of FO order the result is explicitly
\begin{equation}
{\cal H}_{U=\infty}^{\rm FO} =
  +\frac{1}{2} t \sum_{\alpha} \sum_{\langle ij\rangle\parallel\alpha}
  \Big[ 1 + \gamma \sin\psi \cos(\theta - \chi_{\alpha}) \Big]
    f_{i}^{\dagger}f_{j}^{} .
\label{hsfFO}
\end{equation}
Upon Fourier transformation one obtains, reverting to the electron
description, a single band with dispersion depending on the orbital
angles $\{\psi,\theta\}$,
\begin{equation}
\varepsilon_{U=\infty}^{\rm FO}({\bf k})= -t\Big[ A_{\bf k} + \gamma
   \sin\psi \Big( \cos\theta\, C_{\bf k} + \sin\theta D_{\bf k} \Big)
   \Big] ,
\label{dispFO}
\end{equation}
where
\begin{eqnarray}
\label{dispAk}
 A_{\bf k} &=&\cos k_a + \cos k_b + cos k_c ,   \\
\label{dispCk}
 C_{\bf k} &=&\textstyle{\frac{1}{2}}(2\cos k_c-\cos k_a-\cos k_b ), \\
\label{dispDk}
 D_{\bf k} &=&\textstyle{\frac{1}{2}}\sqrt{3} (\cos k_a -\cos k_b ).
\end{eqnarray}
One notes that $C_{\bf k}$ and $D_{\bf k}$ transform as the $\theta$
and $\varepsilon$ components of an $E$ doublet, which makes
Eq.~(\ref{dispFO}) a cubic invariant (i.e., it does not change under the
transformation
$\theta\rightarrow\theta-2\pi/3$ and the simultaneous permutation
$k_a \rightarrow k_b\rightarrow k_c\rightarrow k_a$). It will be useful
to introduce also
\begin{equation}
\label{Pk}
 P_{\bf k} =\sum_{\alpha} e^{+ i \chi_{\alpha}} \cos k_{\alpha}
           = C_{\bf k} + i D_{\bf k},
\end{equation}
which gets multiplied by the phasefactor $e^{-i 2\pi/3}$ under the
permutation $k_a \rightarrow k_b\rightarrow k_c\rightarrow k_a$, as well
as the associated amplitude,
%
%%  Note: Equation below fits to one line, it was cut to two lines
%%        only for the prepint format.
%
\begin{widetext}
%\begin{equation}
%B_{\bf k} = |P_{\bf k}| = \{ C_{\bf k}^2 + D_{\bf k}^2 \}^{1/2}
%   = \{ \cos^2 k_a + \cos^2 k_b + \cos^2 k_c
%     - (\cos k_a \cos k_b + \cos k_b \cos k_c
%              + \cos k_c \cos k_a ) \}^{1/2} ,
%\label{dispBk}
%\end{equation}
\begin{eqnarray}
B_{\bf k} &=& |P_{\bf k}| = \{ C_{\bf k}^2 + D_{\bf k}^2 \}^{1/2} 
   = \{ \cos^2 k_a + \cos^2 k_b + \cos^2 k_c              \nonumber\\
     & & \hskip 3.7cm - (\cos k_a \cos k_b + \cos k_b \cos k_c
                       + \cos k_c \cos k_a ) \}^{1/2} ,
\label{dispBk}
\end{eqnarray}
which transforms as $A_1$, i.e., has cubic symmetry.\cite{noteeta}

Amongst the various phases with AO order let us consider first those of
$G$-type (N\'{e}el-type), denoted by $G$-AO. One obtains from Eq.
(\ref{H_real}) the two-sublattice Hamiltonian ($i\in A$, $j\in B$),
%
%
%%  Note: Equation below fits to one line, it was cut to two lines
%%        only for the prepint format.
%
%\begin{equation}
%{\cal H}_{U=\infty}^{G-{\rm AO}}\!=\!
% \frac{1}{2} t \sum_{\alpha} \sum_{\langle ij\rangle\parallel\alpha}
%  \Big\{\Big[ \cos \psi_{-} \cos \theta_{-}
%           -i \cos \psi_{+} \sin \theta_{-} \Big]
% + \gamma \Big[ \sin \psi_{+} \cos(\theta_{+}\!-\!\chi_{\alpha})
%      +i \sin \psi_{-} \sin(\theta_{+}\!-\!\chi_{\alpha}) \Big]\Big\}
% f_{i}^{\dagger}f_{j}^{} ,
%\label{hsfAO}
%\end{equation}
\begin{eqnarray}
{\cal H}_{U=\infty}^{G-{\rm AO}}\!&=&\!
 \frac{1}{2} t \sum_{\alpha} \sum_{\langle ij\rangle\parallel\alpha}
  \Big\{\Big[ \cos \psi_{-} \cos \theta_{-}
           -i \cos \psi_{+} \sin \theta_{-} \Big]    \nonumber \\
 &+& \gamma \Big[ \sin \psi_{+} \cos(\theta_{+}\!-\!\chi_{\alpha})
      +i \sin \psi_{-} \sin(\theta_{+}\!-\!\chi_{\alpha}) \Big]\Big\}
 f_{i}^{\dagger}f_{j}^{} ,
\label{hsfAO}
\end{eqnarray}
depending on the orbital angles $\{\psi_A,\psi_B,\theta_A,\theta_B\}$,
for which we introduce the shorthand notation for {\it half\/} the
inter-sublattice angles,
\begin{equation}
\psi_{\pm}   = \textstyle{\frac{1}{2}} (\psi_A\pm\psi_B),   \hskip 2cm
\theta_{\pm} = \textstyle{\frac{1}{2}} (\theta_A\pm\theta_B).
\label{anglespm}
\end{equation}
Upon Fourier transformation and diagonalization of the resulting
$2\times 2$ matrix this yields two electron bands (in the reduced
Brillouin zone),
\begin{eqnarray}
\varepsilon_{U=\infty,\pm}^{G-{\rm AO}}({\bf k}) &=& \pm t \Big\{
   \Big[\cos \psi_{-} \cos \theta_{-} A_{\bf k}
     + \gamma \sin \psi_{+}
     \Big( \cos \theta_{+} C_{\bf k} + \sin \theta_{+} D_{\bf k}\Big)
   \Big]^2                      \nonumber \\
 & & + \Big[ \cos \psi_{+} \sin \theta_{-} A_{\bf k}
  - \gamma \sin \psi_{-}
     \Big( \sin \theta_{+} C_{\bf k} - \cos \theta_{+} D_{\bf k}\Big)
   \Big]^2 \Big\}^{1/2}.
\label{dispAO}
\end{eqnarray}
By a similar derivation one may obtain the electronic structure for the
$A$-type and $C$-type AO phases. Using the same notation as above one
finds
\begin{eqnarray}
\label{AAO}
\varepsilon_{U=\infty,\pm}^{A-{\rm AO}}({\bf k})\! &=&\!
t\sum_{\alpha=a,b}\Big\{1+\frac{1}{2}\gamma\big[
 \sin\psi_A\cos(\theta_A-\chi_{\alpha})
+\sin\psi_B\cos(\theta_B-\chi_{\alpha})\big]\Big\}\cos k_{\alpha}
                                                          \nonumber \\
& & \hskip -2.5cm
\pm t\Big\{\big(\frac{\gamma}{2}\sum_{\alpha=a,b}\big[
 \sin\psi_A\cos(\theta_A-\chi_{\alpha})
-\sin\psi_B\cos(\theta_B-\chi_{\alpha})\big]\cos k_{\alpha}\Big)^2
                                                          \nonumber \\
& & \hskip -2.5cm +
 \Big[\Big(\cos\psi_-\cos\theta_- +\gamma\sin\psi_+\cos\theta_+\Big)^2
+\Big(\cos\psi_+\sin\theta_- -\gamma\sin\psi_-\sin\theta_+\Big)^2\Big]
\cos^2k_c\Big\}^{1/2},                                              \\
\label{CAO}
\varepsilon_{U=\infty,\pm}^{C-{\rm AO}}({\bf k})\! &=&\!
t\Big\{\Big(1+\frac{1}{2}\gamma\big[
 \sin\psi_A\cos\theta_A+\sin\psi_B\cos\theta_B\big]\Big)\cos k_c
                                                          \nonumber \\
& & \hskip -2.5cm \pm \Big[\big\{\frac{\gamma}{2}(
 \sin\psi_A\cos\theta_A-\sin\psi_B\cos\theta_B)\cos k_c\big\}^2
                                                          \nonumber \\
& & \hskip -2.5cm
+ \Big\{\cos\psi_-\cos\theta_-(\cos k_a\!+\!\cos k_b)
   +\gamma\sin\psi_+\Big(\cos\big(\theta_+-\frac{2\pi}{3}\big)\cos k_a
         +\cos\big(\theta_++\frac{2\pi}{3}\big)\cos k_b\Big)\Big\}^2
                                                          \nonumber \\
& & \hskip -2.5cm
+ \big\{\cos\psi_+\sin\theta_-(\cos k_a\!+\!\cos k_b)
   -\gamma\sin\psi_-\Big(\sin\big(\theta_+-\frac{2\pi}{3}\big)\cos k_a
   +\sin\big(\theta_++\frac{2\pi}{3}\big)\cos k_b\Big)
   \big\}^2\Big]^{1/2}\Big\}.                             \nonumber \\
\end{eqnarray}
It is now straightforward to derive from Eqs. (\ref{dispFO}) and
(\ref{dispAO})-(\ref{CAO}) the dispersion in any particular
orbital-ordered phase with either {\it complex\/} or {\it real\/}
orbitals.

For the FO$r$ [$\psi=\pi/2$] and the various AO$r$
[$\psi_A=\psi_B=\pi/2$] {\it real states\/} the dispersions are:
\begin{eqnarray}
\label{dispFOr}
\varepsilon_{U=\infty}^{{\rm FO}r}({\bf k}) &=& - t \big[ A_{\bf k}
 \!+\! \gamma \big(\! \cos\theta \, C_{\bf k}
 \!+\! \sin\theta \, D_{\bf k} \big) \big], \\
\label{dispAOr}
\varepsilon_{U=\infty,\pm}^{G-{\rm AO}r}({\bf k}) &=& \pm t
   \Big[ \cos \theta_{-} A_{\bf k}
     + \gamma\Big(\!\cos \theta_{+} C_{\bf k}
                   +\sin \theta_{+} D_{\bf k}\!\Big)\Big],  \\
\label{AAOr}
\varepsilon_{U=\infty,\pm}^{A-{\rm AO}r}({\bf k})& =&
\frac{1}{3}t\Big\{ (2-\gamma \cos\theta_+ \cos\theta_- )
 (A_{\bf k}-C_{\bf k})
+ 3 \gamma \sin\theta_+ \cos\theta_- \; D_{\bf k}         \nonumber \\
&\pm& 
\Big[\Big(\gamma \sin\theta_+ \sin\theta_-(A_{\bf k}-C_{\bf k})
+ 3 \gamma \cos\theta_+ \sin\theta_- \; D_{\bf k}\Big)^2  \nonumber \\
&+&\Big((\cos\theta_-+\gamma\cos\theta_+) (A_{\bf k}+2C_{\bf k})
  \Big)^2 \Big]^{1/2}\Big\},      \\
\label{CAOr}
\varepsilon_{U=\infty,\pm}^{C-{\rm AO}r}({\bf k}) &=&
\frac{1}{3}t\Big\{(1+\gamma \cos\theta_+ \cos\theta_-)
(A_{\bf k}+2C_{\bf k})                                   
\pm \!\Big[\Big( \gamma
 \sin\theta_+ \sin\theta_- (A_{\bf k}+2C_{\bf k}) \Big)^2 \nonumber \\
&+&\Big( \big(2\cos\theta_--\gamma\cos\theta_+\big)
(A_{\bf k}-C_{\bf k})
+3\gamma\sin\theta_+ \; D_{\bf k}\Big)^2\Big]^{1/2}\Big\}.
\end{eqnarray}
\end{widetext}

In contrast to the complex states discussed below, {\it all real
states\/}, whether FO or AO of any type and whatever the value of
$\theta$ (or $\theta_A$ and $\theta_B$), {\it explicitly break cubic
symmetry\/}, i.e., their dispersion is anisotropic. This nonequivalence
between real and complex states is a manifestation of the broken SU(2)
symmetry in the orbital Hubbard model (\ref{H_c}). In extreme cases the
dispersion is two-dimensional (2D). For instance, the dispersion of the
``antiferro'' (i.e. with ${\bm T}_A=-{\bm T}_B$) $G$-type AO state with
alternating $|x\rangle$ and $|z\rangle$ orbitals ($G$-AO$xz$)
[with $\theta_A=0$ and $\theta_B=\pi$],
\begin{equation}
\varepsilon_{U=\infty,\pm}^{G-{\rm AO}xz}({\bf k}) =
   \pm \gamma t D_{\bf k} = \pm \gamma t \frac{\sqrt{3}}{2}
                      \big(\cos k_a -\cos k_b \big),
\label{dispAOzx}
\end{equation}
is 2D because the hopping along the $c$ axis is fully suppressed when
$x^2-y^2$ and $3z^2-r^2$ orbitals alternate. Similarly, the dispersion
of the fully $|x\rangle$-polarized (FO$x$) state ($\theta=\pi$),
\begin{eqnarray}
&&\hskip -1.2cm \varepsilon_{U=\infty}^{{\rm FO}x}({\bf k})
   = - t\big[ A_{\bf k} - \gamma C_{\bf k} \big]        \nonumber \\
&&\hskip -1.2cm = - t\big[ \big(1 + \textstyle\frac{1}{2}\gamma\big)
                \big(\cos k_a + \cos k_b\big)+(1-\gamma)\cos k_c \big],
\label{dispFOx}
\end{eqnarray}
becomes 2D in the genuine orbital case ($\gamma=1$), because when
only $x^2-y^2$ orbitals are occupied, the only type of hopping allowed
in this case, i.e. $\zeta$-hopping [see Eq.~(\ref{H_zetaxi})],
is suppressed along the $c$ axis.

Other states are also anisotropic, but typically have dispersion with
contributions due to all three cubic directions. As an example, the
dispersion of the $G$-type AO state with alternating $3x^2-r^2$ and
$3y^2-r^2$ orbitals ($\theta_A=-\theta_B=2\pi/3$) along the $a$ and $b$
cubic axes ($G$-AO$ab$),
\begin{eqnarray}
&&\hskip -1.2cm \varepsilon_{U=\infty,\pm}^{G-{\rm AO}ab}({\bf k})
  =\pm t\big[ -\textstyle{\frac{1}{2}} A_{\bf k}+\gamma C_{\bf k}\big]
                                                            \nonumber \\
  && \hskip -1.2cm =\mp
  \textstyle{\frac{1}{2}}t\big[(1\!+\!\gamma) (\cos k_a + \cos k_b)
      + \big(1-2\gamma\big)\cos k_c\big],
\label{dispAOab}
\end{eqnarray}
is cubic at $\gamma=0$, but becomes predominantly but not fully 2D for
$\gamma=1$. By contrast, the dispersion of the
($\theta_A=-\theta_B=\pi/2$) state, with alternation between symmetric
and antisymmetric combinations, $(|x\rangle+|z\rangle)$ and
$(|x\rangle-|z\rangle)$, called $G$-AO$sa$,
\begin{eqnarray}
\varepsilon_{U=\infty,\pm}^{G-{\rm AO}sa}({\bf k})
   &=& \pm \gamma t C_{\bf k}                               \nonumber \\
  = \pm \gamma t && \hskip -0.7cm
     \big[ -\textstyle{\frac{1}{2}} (\cos k_a + \cos k_b)
       + \cos k_c \big],
\label{dispAOxpmz}
\end{eqnarray}
is quasi-one-dimensional (quasi-1D), qualitatively similar to that of
the $|z\rangle$-polarized (FO$z$) state ($\theta=0$),
\begin{eqnarray}
&&\hskip -1.2cm \varepsilon_{U=\infty}^{{\rm FO}z}({\bf k})
   = - t \big[ A_{\bf k} + \gamma C_{\bf k} \big]           \nonumber \\
&& \hskip -1.2cm =- t\big[
   \big(1 -\textstyle\frac{1}{2}\gamma\big)\big(\cos k_a + \cos k_b\big)
             +(1+\gamma)\cos k_c\big],
\label{dispFOz}
\end{eqnarray}
which becomes quasi-1D in the orbital case ($\gamma=1$).

\begin{figure}[t!]
\includegraphics[width=7.7cm]{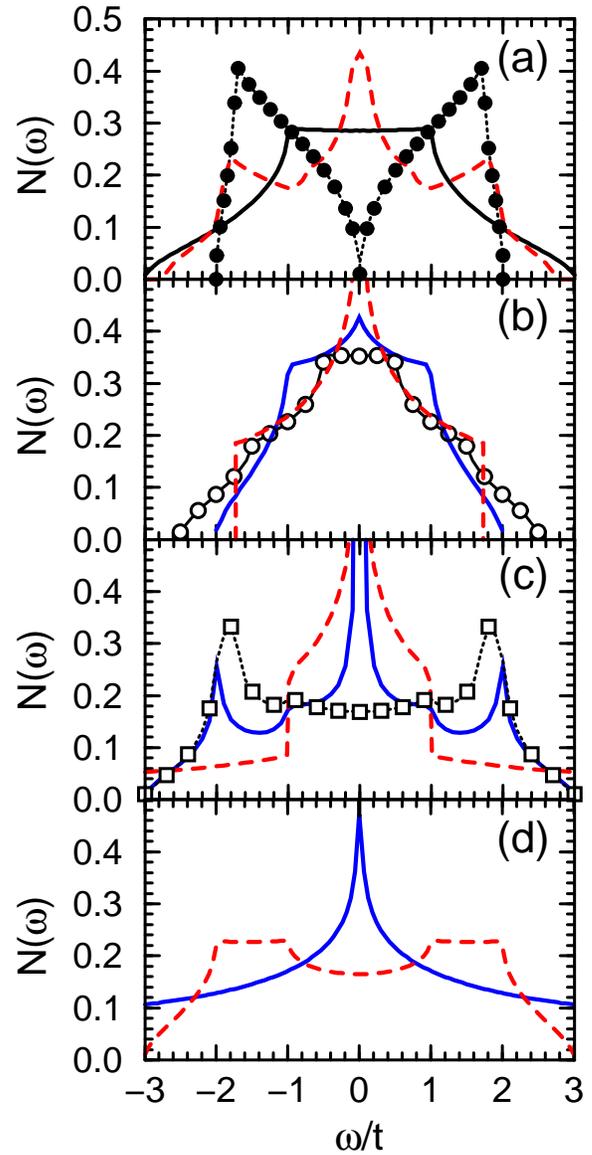}
\caption{
(Color online) Density of states $N(\omega)$ at $\gamma=1$
(in units of $t=1$) for different orbital ordered phases:
(a) complex orbital order: uniform FO+ [degenerate with $A$-AO$\pm$]
    (solid line), $G$-AO$\pm$ (filled circles), and $C$-AO$\pm$
    (dashed line);
(b) alternating real orbital order in $G$-type phases:
    $G$-AO$sa$ (solid line) [the same density of states is obtained for
    $C$-AO$sa$ phase], $G$-AO$xz$ (dashed line),
    and $G$-AO$ab$ (circles);
(c) alternating real orbital order in selected $C$- and $A$-type phases:
    $C$-AO$xz$ (solid line), $A$-AO$xz$ (dashed line),
    and $A$-AO$sa$ (squares);
(d) real uniform orbital order FO$x$ (solid line)
                           and FO$z$ (dashed line).
}
\label{fig:dos}
\end{figure}

\begin{figure}[t!]
\includegraphics[width=7.9cm]{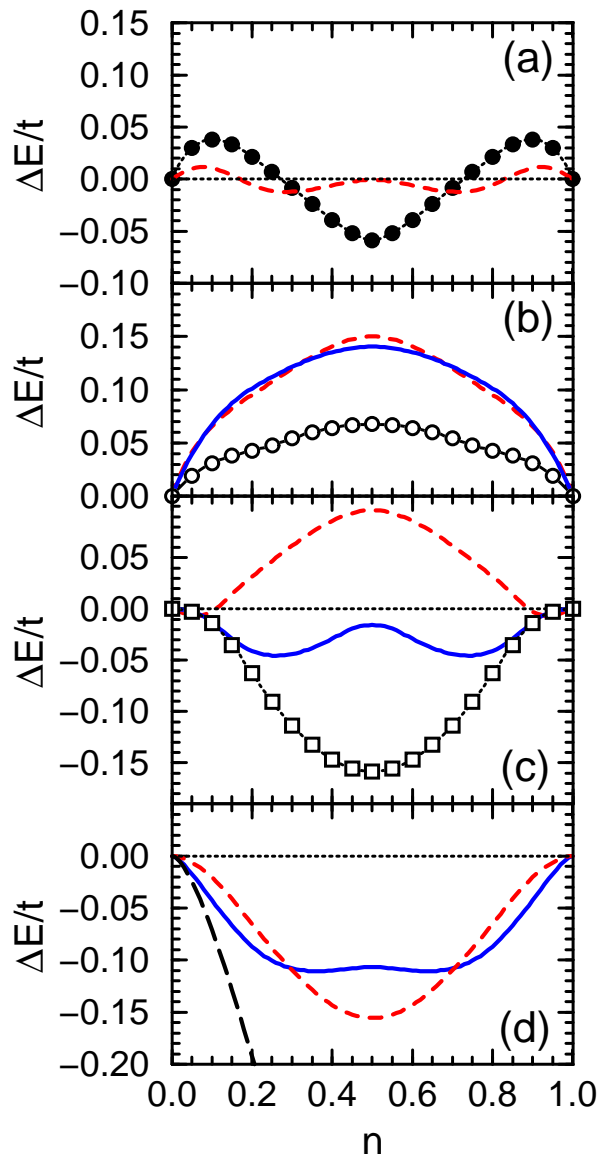}
\caption{
(Color online) Kinetic energy gain (loss) $\Delta E/t$ with
respect to FO+ phase [reference zero energy given by horizontal dotted
lines], as a function of electron filling $n$ for various orbital
ordered phases (at $U=\infty$):
(a) complex orbital order:
    $G$-AO$\pm$ (filled circles), and $C$-AO$\pm$ (dashed line);
(b) alternating real orbital order in $G$-type phases:
    $G$-AO$sa$ (solid line) [degenerate with $C$-AO$sa$ phase],
    $G$-AO$xz$ (dashed line), and $G$-AO$ab$ (circles);
(c) alternating real orbital order in selected $C$- and $A$-type phases:
    $C$-AO$xz$ (solid line), $A$-AO$xz$ (dashed line),
    and $A$-AO$sa$ (squares);
(d) uniform real orbital order FO$x$ (solid line)
                           and FO$z$ (dashed line).
The long-dashed line in (d) shows the kinetic energy for noninteracting
electrons in the $e_g$ band (disordered phase at $U=0$).
}
\label{fig:Ekin}
\end{figure}

The reduced symmetry of the FO$x$ and FO$z$ states is reflected in their
respective densities of states, shown in Fig. \ref{fig:dos}(d), which
lead to favorable kinetic energies [see Fig. \ref{fig:Ekin}(d)], as
discussed in Sec. \ref{sec:dos}.
Obviously, such broken-symmetry states could be favored either in low
dimensional systems, as the FO$x$ state found for a 2D square
lattice,\cite{Mac99} and suggested for bilayer manganites,\cite{Maebi}
or by a strong JT effect favoring a particular type of occupied $e_g$
orbitals due to oxygen distortions, as realized for instance in bilayer
systems.\cite{Koi01} The latter applies also for the $G$-type AO states,
which have typically smaller bandwidths than the FO states; a few
examples are shown in Figs. \ref{fig:dos}(b) and \ref{fig:dos}(c).

As illustrative examples of the $A$-type and $C$-type phases with
alternating real orbitals [either along the $c$ axis or in the $(a,b)$
planes], we give dispersions in each case for:
 (i) $\theta_A=-\theta_B=\frac{\pi}{2}$, i.e. with alternating
     $(|x\rangle\pm |z\rangle)/\sqrt{2}$ states, and
(ii) $\theta_A=0$, $\theta_B=\pi$, with alternating $|x\rangle$ and
     $|z\rangle$ states,
\begin{widetext}
\begin{eqnarray}
\label{AAOsa}
\varepsilon_{U=\infty,\pm}^{A-{\rm AO}sa}({\bf k})
  &=& \frac{1}{3}t\Big\{2 (A_{\bf k}-C_{\bf k})\pm\gamma
\big[(A_{\bf k}+2C_{\bf k})^2+9D_{\bf k}^2\big]^{1/2}\Big\} \nonumber \\
  &=& t\Big\{\cos k_a+\cos k_b \pm \gamma \Big[
   \frac{3}{4}(\cos k_a\!-\!\cos k_b)^2
                       +\cos^2 k_c\Big]^{1/2}\Big\},                  \\
\label{AAOxz}
\varepsilon_{U=\infty,\pm}^{A-{\rm AO}xz}({\bf k})
&=& \frac{1}{3}t\big(2\pm\gamma\big) (A_{\bf k}-C_{\bf k})
 =  t\Big(1\pm\frac{1}{2}\gamma\Big)(\cos k_a+\cos k_b),              \\
\label{CAOsa}
\varepsilon_{U=\infty,\pm}^{C-{\rm AO}sa}({\bf k})
   &=& \frac{1}{3}t\Big[ (1\pm\gamma) A_{\bf k}
   +(2\mp\gamma) C_{\bf k}\Big]
   =  t\Big[\cos k_c\pm\frac{1}{2}\gamma(\cos k_a+\cos k_b)\Big],     \\
\label{CAOxz}
\varepsilon_{U=\infty,\pm}^{C-{\rm AO}xz}({\bf k})
   &=& \frac{1}{3}t\Big\{A_{\bf k}+2C_{\bf k}\pm\gamma\big[
  (A_{\bf k}+2C_{\bf k})^2+9D_{\bf k}^2\big]^{1/2}\Big\}    \nonumber \\
   &=& t \Big[\cos k_c \pm \gamma \Big\{\frac{3}{4}
 (\cos k_a\!-\!\cos k_b)^2+\cos^2 k_c\Big\}^{1/2}\Big].
\end{eqnarray}
\end{widetext}
The anisotropy of these phases is quite strong, and the $A$-AO$xz$ phase
has even a 2D dispersion.

Finally we consider the orbital ordered states with complex orbitals.
For two of these {\it complex states\/}, namely the ferro
$|+\rangle$-polarized orbital order (FO+) [$\psi=0$] and the $G$-type
alternating orbital order (AO$\pm$) with $|+\rangle/|-\rangle$ staggered
orbitals [with $\psi_A=0$ and $\psi_B=\pi$], all cubic directions are
equivalent, and one finds the dispersions
\begin{equation}
\varepsilon_{U=\infty}^{{\rm FO+}}({\bf k})= - t A_{\bf k} ,
\label{dispFO+}
\end{equation}
and
\begin{equation}
\varepsilon_{U=\infty,\pm}^{G-{\rm AO\pm}}({\bf k}) =
    \pm \gamma t B_{\bf k} ,
\label{dispAOpm}
\end{equation}
respectively. Thus, one finds that the dispersion of the FO+ state and
its density of states, shown in Fig.~\ref{fig:dos}(a), is that of a
simple cubic lattice, as it originates entirely from the
pseudospin-conserving hopping $\propto c_{i\pm}^{\dagger}c_{j\pm}^{}$,
because at $U \to \infty$ the alternating, non-pseudospin-conserving,
hopping is fully suppressed by the imposed FO+ order. The reverse is
true in the $G$-AO$\pm$ state: here the dispersion $\propto \pm B_{\bf
k}$ comes entirely from the alternating hopping
$\propto c_{i\pm}^{\dagger}c_{j\mp}^{}$, as the pseudospin-conserving
hopping is fully suppressed by the AO$\pm$ order. It is an important
feature of both these complex states, built from cubic orbitals, that
they {\it retain cubic symmetry\/}. Precisely for that reason these
complex orbital ordered states were proposed as candidates for the
ground state of the FM metallic phase of the manganites,
\cite{Bri01,Shi00} to explain the observed cubic symmetry of the magnon
spectra.\cite{Per96,Fer98}

%%%%%%%%%%%%%%%%%%%%%%%%%%%%%%%%%%%%%%%%%%%%%%%%%%%%%%%%%%%%%%%%%%%%%%%%
%%
%%                             figure 2
%%
%%%%%%%%%%%%%%%%%%%%%%%%%%%%%%%%%%%%%%%%%%%%%%%%%%%%%%%%%%%%%%%%%%%%%%%%

%%%%%%%%%%%%%%%%%%%%%%%%%%%%%%%%%%%%%%%%%%%%%%%%%%%%%%%%%%%%%%%%%%%%%%%%
%%
%%                             figure 3
%%
%%%%%%%%%%%%%%%%%%%%%%%%%%%%%%%%%%%%%%%%%%%%%%%%%%%%%%%%%%%%%%%%%%%%%%%%

The other two orbital-ordered complex states break explicitly cubic
symmetry: the $A$-type and $C$-type AO$\pm$ states. In the $A$-AO$\pm$
state layers of $|+\rangle$ and $|-\rangle$ orbitals alternate in the
$c$ direction, resulting in the dispersion
\begin{eqnarray}
& & \hskip -1.2cm
\varepsilon_{U=\infty,\pm}^{A-{\rm AO\pm}}({\bf k})
 = \frac{1}{3}t\big[(2\pm \gamma) A_{\bf k}
                     -2(1 \mp \gamma) C_{\bf k} \big]
                     \nonumber \\
& & \hskip 0.6cm
 = t\big( \cos k_a+\cos k_b \pm \gamma \cos k_c  \big).
\label{AAOpm}
\end{eqnarray}
This dispersion is qualitatively equivalent to that of the FO+ state,
and thus the densities of states of the FO+ and $A$-AO$\pm$ phases are
the same.
The reason is that replacing in the $c$-direction every second
$|+\rangle$ orbital by a $|-\rangle$ orbital does not affect the hopping
parameter along $c$, and so the resulting doubling of the unit cell only
halves the Brillouin zone without changing the dispersion.
In the $C$-AO$\pm$ state instead columns of $|+\rangle$ and $|-\rangle$
orbitals alternate in the $(a,b)$ planes, and one finds
\begin{eqnarray}
\varepsilon_{U=\infty,\pm}^{C-{\rm AO\pm}}({\bf k})
 &=& \frac{1}{3}t\Big\{A_{\bf k}+2C_{\bf k}\pm \gamma
     \Big[9 D_{\bf k}^2+ (A_{\bf k}-C_{\bf k})^2\Big]\Big\}
                 \nonumber \\
   & & \hskip -2.5cm = t \big[\cos k_c \pm \gamma
      \big(\cos^2 k_a\!+\!\cos^2 k_b - \cos k_a \cos k_b \big)^{1/2}
                \big].
\label{CAOpm}
\end{eqnarray}
In contrast to the FO+, $G$-AO$\pm$ and $A$-AO$\pm$ phases, the
$C$-AO$\pm$ phase is not cubic symmetric.

The densities of states of the complex states show a gradual
crossover with increasing alternating orbital character from the full
bandwidth of $6t$ for the FO+ and $A$-AO$\pm$ phases, obtained also at
$U=0$ both for the spin problem and for
the $e_g$ band (Fig. \ref{fig:dosg}), to a narrower bandwidth of
$2(1+\sqrt{3})t$ for the $C$-AO$\pm$ phase, and finally to a bandwidth
of $4t$ for the $G$-AO$\pm$ phase. It is remarkable that, upon going
from the FO+ phase to the $A$-AO$\pm$ phase, the change
from uniform to alternating orbital order along only one cubic direction
does not modify the density of states, while the density of states
changes its shape completely upon going to the $G$-AO$\pm$ phase, with
a large spectral weight accumulated now close to the band edges,
resulting in a quite peculiar density of states with large maxima close
to $|\omega|\simeq 2\gamma t$, separated by a minimum with $N(0)=0$ at
$\omega=0$ [Fig. \ref{fig:dos}(a)].
The density of states for the $C$-AO$\pm$ phase has a width of
$2(1+\sqrt{3})t$, and represents an intermediate case, having
some features in common with that of the $G$-AO$\pm$ phase.

\subsection{Densities of states and kinetic energies
            in orbital ordered states}
\label{sec:dos}

It is worthwile to consider next the densities of states of various
orbital ordered states in a little more detail (see Fig. \ref{fig:dos}),
and investigate their consequences for the kinetic energy (Fig.
\ref{fig:Ekin}).
Focusing first on the bandwidth, we note that for any FO state this
takes the maximum attainable value $6t$. This result is not limited to
the FO states considered explicitly above, i.e., the complex FO+ and the
real FO$z$ and FO$x$ states, for which it was already pointed out by Van
den Brink and Khomskii,\cite{Bri01} but holds in general, i.e., for
arbitrary $\psi$ and $\theta$, as readily shown from Eq.~(\ref{dispFO}).
Moreover, the result is independent of $\gamma$ and so holds both in the
orbital case and in the spin case.

By contrast, in any $G$-type AO state the bandwidth is {\it smaller\/}
than $6t$ [see Fig. \ref{fig:dos}(b)], and depends on $\gamma$. In
particular, in any $G$-type ``antiferro'' state (with
${\bm T}_A=-{\bm T}_B$, so $\psi_B=\pi-\psi_A$ and
$\theta_B=\theta_A-\pi$), such as the complex $G$-AO$\pm$ state or the
real $G$-AO$xz$ and $G$-AO$sa$ states considered above, the width is
proportional to $\gamma$ (viz. $4\gamma t$, $2\sqrt{3}\gamma t$, and
$4\gamma t$, respectively for those three) as follows from Eq.
(\ref{dispAO}). In such a state the bandwidth therefore, correctly,
collapses to zero in the spin case ($\gamma=0$) where hopping is
completely suppressed by the AF spin order.\cite{noteDE} The important
point to note here is that at finite $\gamma$, and so in particular in
the genuine orbital case, the bandwidth even of an ``antiferro'' state
is {\it finite\/} though smaller than that of the FO states. Thus, while
in the spin case the kinetic energy of carriers is fully lost when going
from FM to N\'{e}el-type AF order, this is not so for the analogous FO
to $G$-type AO transition in the orbital case.

One might still be tempted to believe that, as familiar from the spin
case, also in the orbital case FO order is most favorable for lowering
the kinetic energy of charge carriers, simply because this gives the
largest bandwidth. However, the situation is not that simple, not only
because there are several inequivalent FO states with different
densities of states which have nevertheless the same bandwidth, but also
because some $A$-AO and $C$-AO phases have again the same bandwidth,
and so one really has to consider the details of the density of states
in each case. This is demonstrated in Fig.~\ref{fig:Ekin}, which shows
the kinetic energy gain $\Delta E$ with respect to the complex FO+ state
as a function of electron filling $n$ for various FO and AO states with
(complex or real) orbital order, obtained by straightforward integration
of the respective density of states. Indeed, at small electron filling
$n$, and also at small doping $x=1-n$, $\Delta E$ is lower for the (FO,
$A$-AO and $C$-AO) states with full bandwidth $6t$ than for any state
with a narrower density of states, in particular for the $G$-type AO
states of Fig. \ref{fig:Ekin}(b), because the first doped holes enter in
the former case with an energy $\sim -3t$ close to the band edge, while
the lowest accessible energy is higher in all $G$-AO states. Note that
the orbital order observed in LaMnO$_3$ is close to that of the
$C$-AO$sa$ phase,\cite{noteMang} and this phase has the same density of
states as the $G$-AO$sa$ phase [see Fig. \ref{fig:dos}(b)], and thus has
a rather unfavourable kinetic energy [Fig.~\ref{fig:Ekin}(b)].
This demonstrates that both an interplay between spin and orbital order
due to the SE interactions at finite $U$, and the JT interactions
between orbitals on neighboring sites, induced by the coupling to the
lattice, play an important role in real materials and stabilize the
orbital order observed in undoped LaMnO$_3$.\cite{Fei99,Feh04}

Among the states with AO order of real orbitals, but FO order along one
or two cubic directions, we identified three phases, $C$-AO$xz$,
$A$-AO$xz$, and $A$-AO$sa$, which have lower energies than the FO+ phase
close to $n=0$ and $n=1$ [Fig.~\ref{fig:Ekin}(c)]. All of them have the
full bandwidth $6t$ [Fig.~\ref{fig:dos}(c)], but a finite density of
states at $\omega=-3t$ gives the $A$-AO$xz$ phase the lowest energy of
these phases at very low $n$ or $x$. At somewhat higher filling
$n\sim 0.07$ (doping $x\sim 0.07$) the other two phases take over, and
are in fact more stable than the FO+ phase in the entire regime of $n$.
This follows from the large densities of states of these phases at
$|\omega|\simeq 2t$. In contrast, the $A$-AO$xz$ phase with a large
spectral weight close to $\omega=0$ has a higher energy than the FO+
phase in a broad range of electron filling $0.11<n<0.89$.

The above discussion shows that at finite but still rather modest
electron filling or doping, the overall shape of the density of states
becomes more important, and the states with large density of states
near the band edges could be favored {\it a priori\/}, even in cases
when the bandwidth is smaller that $6t$. An interesting example here
is the complex ``antiferro'' $G$-AO$\pm$ state, with its energy falling
below that of the complex FO+ state for $n>0.27$ or $x>0.27$ [Fig.
\ref{fig:Ekin}(a)], because of the large number of states available
in the $G$-AO$\pm$ state just close to the band edges at $|\omega|=2t$,
whereas in the FO+ state the energy of available electron states,
though initially $-3t$, rises rapidly with increasing doping [Fig.
\ref{fig:dos}(a)]. However, in reality the transition from FO+ to
$G$-AO$\pm$ state does not happen, as the real FO (FO$z$ and FO$x$)
states have even lower kinetic energy throughout than both complex
states. This can be ascribed to the lower-dimensional nature of their
dispersion and the resulting different location of the Van Hove
singularities, which [compare Fig.~\ref{fig:dos}(d)] enhances the
density of states near the band edges at $\pm 3t$ and at the band
center for the 2D FO$x$ state, and in the intermediate range
$t\lesssim |\omega|\lesssim 2t$ for the quasi-1D FO$z$ state. As a
result, at small filling (doping) the kinetic energy gain $\Delta E$
is the lowest one for the FO$x$ state, while at larger filling
$n\gtrsim 0.30$ (doping $x\gtrsim 0.30$), the FO$z$ state takes over.
However, in this regime of electron filling the energy gain $\Delta E$
for the $A$-AO$sa$ phase is lower by a few percent, and the two phases
may be considered as practically degenerate.

For comparison and later reference we have included in
Fig.~\ref{fig:Ekin} also the kinetic energy for the uncorrelated $e_g$
band (the correlated OL phase is analyzed in Sec. \ref{sec:oliq}).
Of course,
at $U=0$ any kind of orbital order is absent and one finds by far
the lowest kinetic energy for the disordered $e_g$ orbitals.
The $e_g$ bands have then the dispersion given by
\begin{equation}
\varepsilon_{U=0,\pm}({\bf k})= - t ( A_{\bf k} \pm \gamma B_{\bf k} ).
\label{disp_unc}
\end{equation}
Remarkably, these bands at $U=0$ represent formally a superposition of
the FO+ and $G$-AO$\pm$ bands at $U=\infty$,
\begin{equation}
\varepsilon_{U=0,\pm}({\bf k})=
    \varepsilon_{U=\infty}^{\rm FO+}({\bf k})
     + \varepsilon_{U=\infty,\pm}^{G-{\rm AO\pm}}({\bf k}) .
\label{disp_superpos}
\end{equation}
and so naturally also show full cubic symmetry and a bandwidth equal to
$6t$ [Fig. \ref{fig:dosg}(d)]. One notes that, because both
pseudospin-conserving and non-pseudospin-conserving hopping channels
fully contribute here, considerably more kinetic energy can be gained
than in any of the orbital-ordered states. In particular, as Fig.
\ref{fig:dosg}(d) shows, there is a large density of states at and near
the band edges, and thus $\Delta E$ is the lowest in this disordered
state already at small electron filling $n$, and then remains so
throughout. Of course, this large kinetic energy gain will be partly
lost for large $U$ near $n=1$, where at least one hopping channel gets
partially suppressed by electron correlations. However, the result here
indicates that the tendency towards the OL state with disordered $e_g$
orbitals is particularly pronounced. We shall come back to this point,
presenting more evidence in favor of the correlated OL phase, in Sec.
\ref{sec:oliq}.

%%%%%%%%%%%%%%%%%%%%%%%%%%%%%%%%%%%%%%%%%%%%%%%%%%%%%%%%%%%%%%%%%%%%%%%%
%%
%%                      Hartree-Fock approximation
%%
%%%%%%%%%%%%%%%%%%%%%%%%%%%%%%%%%%%%%%%%%%%%%%%%%%%%%%%%%%%%%%%%%%%%%%%%
\section{Hartree-Fock approximation}
\label{sec:hf}

\subsection{Instabilities towards orbital order}
\label{sec:hfgen}

We turn now to the orbital Hubbard model (\ref{H_c}) with finite $U$,
where it is to be expected that polarization, when it occurs, need not
be complete but can be partial, as in the spin case. Also, the
existence of orbital ordered states will in general require a
sufficiently large $U/t$. Which instabilities towards orbital ordering
occur and at what value of $U/t$ can be investigated either by
considering the corresponding susceptibilities, e.g. in random phase
approximation,\cite{Shi00} or by comparing the energies determined in
the HF approximation.\cite{Bri01} If various ordered states are
possible, one needs to calculate their energy (or free energy at
finite temperature) to determine which one is actually realized.

\begin{widetext}
In the absence of SU(2) symmetry it is not sufficient to decouple the
interaction term in Eq. (\ref{H_c}) in the familiar mean-field way,
$n_{i+}n_{i-}\simeq
(\langle n_{i+}\rangle n_{i-}+n_{i+}\langle n_{i-}\rangle
-\langle n_{i+}\rangle\langle n_{i-}\rangle)$,
but one needs instead the general HF decoupling,
\begin{equation}
 n_{i+}n_{i-} \simeq \Big( \langle n_{i+}\rangle n_{i-}
 +n_{i+}\langle n_{i-}\rangle
 -\langle n_{i+}\rangle\langle n_{i-}\rangle \Big)
 - \left(\langle T_i^+ \rangle c_{i-}^{\dagger}c^{}_{i+}
        +c_{i+}^{\dagger}c^{}_{i-}\langle T_i^-\rangle
        -\langle T_i^+\rangle \langle T_i^-\rangle \right).
\label{HFdec}
\end{equation}
In the FO case, i.e., when one assumes a single three-component order
parameter,
$T_z=\langle T_i^z\rangle$,
$T_+=\langle T_i^+\rangle$,
$T_-=\langle T_i^-\rangle$,
one obtains upon Fourier transformation the HF Hamiltonian
\begin{eqnarray}
H_{\rm HF}^{\rm FO} &=& \sum_{\bf k}
  \left( c_{{\bf k}+}^{\dagger} \; c_{{\bf k}-}^{\dagger} \right)
\left( \begin{array}{cc}
\textstyle{\frac{1}{2}} U n - U T_z - t A_{\bf k} &
- U T_- - \gamma t P_{\bf k}^{\ast}
   \\[0.2cm]
- U T_+ - \gamma t P_{\bf k}^{}
&
\textstyle{\frac{1}{2}} U n + U T_z - t A_{\bf k}
\end{array}
\right)
  \left(\begin{array}{c} c_{{\bf k}+}^{} \\[0.2cm]
                      c_{{\bf k}-}^{} \end{array} \right) \nonumber \\
 & &  - \textstyle{\frac {1}{4}} U n^2 + U ( T_z^2 + T_+ T_- )  ,
 \label{HamHFFO}
\end{eqnarray}
with $P_{\bf k}$ given by Eq.~(\ref{Pk}).
The eigenvalues are (with $T_+ = T e^{i \theta}$,
$T_- = T e^{-i \theta}$, so that $T_+ T_- = T^2 = T_x^2 + T_y^2$)
\begin{equation}
\varepsilon_{\pm}^{\rm FO}({\bf k})=-t A_{\bf k}
     + U\big(\textstyle{\frac{1}{2}}n \pm \hat{E}_{\bf k}\big),
\label{HFevFO}
\end{equation}
where
\begin{equation}
\hat{E}_{\bf k} =\Big[ T^2 + T_z^2
  + 2 \frac{\gamma t}{U} T
           \big( \cos \theta C_{\bf k} + \sin \theta D_{\bf k} \big)
  + \Big(\frac{\gamma t}{U}\Big)^2B_{\bf k}^2\Big]^{1/2},
\label{Ehatk}
\end{equation}
and the HF groundstate energy per site is then given by
\begin{equation}
E_{\rm HF}^{\rm FO} = \frac{1}{N} \sum_{\bf k}
   \Big[ n_{-}({\bf k}) \varepsilon_{-}^{\rm FO}({\bf k})
   + n_{+}({\bf k}) \varepsilon_{+}^{\rm FO}({\bf k})\Big]
   - \frac{1}{4} U n^2 + U \big( T^2 + T_z^2 \big),
\label{HFenergyFO}
\end{equation}
\end{widetext}
where $n_{-}({\bf k})$ [$n_{+}({\bf k})$] is the occupation number of
the lower (upper) band. For large $U$ ($\gtrsim 6t$) a gap opens,
and so for less than half-filling only the lower band is occupied.
Setting the derivatives of $E_{\rm HF}^{\rm FO}$ with respect to $n$,
$T_z$, $T$, and $\theta$ equal to zero then yields the self-consistency
equations
\begin{eqnarray}
\hskip -.7cm
n &=& \frac{1}{N}\sum_{\bf k} n_{-}({\bf k}) ,
        \label{nHF} \\
\hskip -.7cm
T_z &=&  \frac{1}{2}\frac{1}{N}\sum_{\bf k} n_{-}({\bf k})
   \frac{T_z}{\hat{E}_{\bf k}}  ,
        \label{TzHF} \\
\hskip -.7cm
T &=&  \frac{1}{2}\frac{1}{N}\sum_{\bf k} n_{-}({\bf k})
   \frac{ T + \frac{\gamma t}{U}
        \big( \cos\theta\, C_{\bf k} + \sin\theta\, D_{\bf k} \big) }
                {\hat{E}_{\bf k}}  ,
        \label{THF} \\
\hskip -.7cm
0 &=&  \frac{1}{2N}\sum_{\bf k} n_{-}({\bf k})
   \frac{T\big(\sin\theta\, C_{\bf k} - \cos\theta\, D_{\bf k}\big) }
                {\hat{E}_{\bf k}}  .
\label{thetaHF}
\end{eqnarray}
While Eq.~(\ref{nHF}) is trivially satisfied in the sense that it
simply fixes the Fermi level for given filling $n$, two general
conclusions can be proven from the remaining three equations. Firstly,
it follows from Eq.~(\ref{thetaHF}), because of the dependence of
$\hat{E}_{\bf k}$ on $C_{\bf k}$ and $D_{\bf k}$ [see Eq.~(\ref{Ehatk})]
and the explicit form of the latter two functions [see Eqs.
(\ref{dispCk}) and (\ref{dispDk})], that for nonzero $T$ the azimuth
$\theta$ must equal either $0$ (or equivalently $+2\pi/3$ or $-2\pi/3$)
or $\pi$ (or equivalently $-\pi/3$ or $+\pi/3$), i.e. the projection of
the order parameter on the `real' equatorial plane has to be along one
of the cubic directions. Secondly, it follows that both a purely
complex state (i.e., $T=0$, $T_z \neq 0$) and a purely real state
(i.e., $T_z=0$, $T \neq 0$) are permissible states, in the sense that
$T_z=0$ is a self-consistent solution of Eq. (\ref{TzHF}) and
alternatively $T=0$ is one of Eq.~(\ref{THF}). We remark that both these
properties of the possible states need not be postulated or assumed but
are proven here from the HF self-consistency equations.

As $C$-type and $A$-type AO phases would give qualitatively similar
results, we will consider from now on only $G$-type AO phases, and
denote them for brevity by ``AO'' instead of by ``$G$-AO''. So we assume
independent three-component order parameters on interlacing sublattices
A and B,
$T_z^{\rm A}=\langle T_i^z\rangle_{\rm A}$,
$T_z^{\rm B}=\langle T_i^z\rangle_{\rm B}$, etc.
Then the HF Hamiltonian is
\begin{widetext}
\begin{eqnarray}
H_{\rm HF}^{\rm AO} &=& \sum_{\bf k}
  \left( \begin{array}{c}
         c_{{\rm A},{\bf k}+}^{\dagger} \\
         c_{{\rm A},{\bf k}-}^{\dagger} \\
         c_{{\rm B},{\bf k}+}^{\dagger} \\
         c_{{\rm B},{\bf k}-}^{\dagger} \end{array} \right)^{\!\!\rm T}
         \nonumber
\left( \begin{array}{cccc}
\textstyle{\frac{1}{2}} U n - U T_z^{\rm A} & - U T_-^{\rm A} &
  - t A_{\bf k} & - \gamma t P_{\bf k}^{\ast}
   \\[0.2cm]
- U T_+^{\rm A} & \textstyle{\frac{1}{2}} U n + U T_z^{\rm A} &
 - \gamma t P_{\bf k} & - t A_{\bf k}
   \\[0.2cm]
- t A_{\bf k} & - \gamma t P_{\bf k}^{\ast} &
  \textstyle{\frac{1}{2}} U n - U T_z^{\rm B}  & - U T_-^{\rm B}
   \\[0.2cm]
- \gamma t P_{\bf k} & - t A_{\bf k} &
  - U T_+^{\rm B} & \textstyle{\frac{1}{2}} U n + U T_z^{\rm B}
\end{array}
\right)
  \left( \begin{array}{c}
        c_{{\rm A},{\bf k}+}^{} \\[0.2cm]
        c_{{\rm A},{\bf k}-}^{} \\[0.2cm]
        c_{{\rm B},{\bf k}+}^{} \\[0.2cm]
        c_{{\rm B},{\bf k}-}^{} \end{array} \right)   \nonumber \\
 & &  - \textstyle{\frac {1}{4}} U n^2
  +\textstyle{\frac{1}{2}} U\big[
          \big(T_z^{\rm A}\big)^2 + T_+^{\rm A} T_-^{\rm A}
        + \big(T_z^{\rm B}\big)^2 + T_+^{\rm B} T_-^{\rm B} \big] .
 \label{HamHFAO}
\end{eqnarray}
Like above, the HF groundstate energy per site is then formally given
(with $T_+^{\rm A} = T^{\rm A} e^{i \theta_{\rm A}}$, etc.) by
\begin{equation}
E_{\rm HF}^{\rm AO} = \frac{1}{N} \sum_{\beta=1}^4 \sum_{\bf k}
   n_{\beta}({\bf k}) \varepsilon_{\beta}^{\rm AO}({\bf k})
   - \textstyle{\frac {1}{4}} U n^2
   + \textstyle{\frac {1}{2}} U \big[
                \big(T^{\rm A}\big)^2 + \big(T_z^{\rm A}\big)^2
              + \big(T^{\rm B}\big)^2 + \big(T_z^{\rm B}\big)^2 \big] ,
\label{HFenergyAO}
\end{equation}
\end{widetext}
where the sum on $\beta$ is over the four bands and that on ${\bf k}$
is over the reduced Brillouin zone. However, as the $4 \times 4$ matrix
in Eq.~(\ref{HFenergyAO}) cannot be diagonalized analytically in the
general case (i.e. for arbitrary order parameters), no further progress
can be made like in the FO case. In particular one cannot strictly prove
that purely real or purely complex states are permissible solutions.

Yet this still seems likely, and if one makes this assumption, then for
the case of the complex (AO$c$) state, i.e. with
$T^{\rm A}=T^{\rm B}=0$, the $4 \times 4$ matrix simplifies enough to
obtain explicit expressions for the band dispersions,
\begin{equation}
\varepsilon_{\beta}^{\rm AOc}({\bf k})=
     + U\big(\textstyle{\frac{1}{2}}n \pm \hat{F}_{1,{\bf k}}
                                  \pm \hat{F}_{2,{\bf k}} \big),
\label{HFevAOpm}
\end{equation}
where
\begin{eqnarray}
\label{F1hat}
\hat{F}_{1,{\bf k}} &=& \Big[
      \Big(\frac{ T_z^{\rm A} + T_z^{\rm B} }{2}\Big)^2
    + \Big(\frac{\gamma t}{U}\Big)^2 B_{\bf k}^2\Big]^{1/2},     \\
\label{F2hat}
\hat{F}_{2,{\bf k}} &=& \Big[
      \Big(\frac{ T_z^{\rm A} - T_z^{\rm B} }{2}\Big)^2
    + \Big(\frac{t}{U}\Big)^2 A_{\bf k}^2\Big]^{1/2}.
\end{eqnarray}
Setting the derivatives of $E_{\rm HF}^{\rm AOc}$ with respect to $n$,
$T_z^{\rm A}$ and $T_z^{\rm B}$ to zero yields again HF self-consistency
equations. From these one easily proves that $T_z^{\rm A}=-T_z^{\rm B}$,
i.e. that the stable complex state is actually the AO$\pm$ state.

For the case of a real (AOr) state, i.e., with
$T_z^{\rm A}=T_z^{\rm B}=0$, an analytic solution is also possible,
but this is so unwieldy as to be impractical. However, if one further
assumes that $T^{\rm A}=T^{\rm B}\equiv T$ one can derive the
approximate expressions
\begin{eqnarray}
\varepsilon_{\beta}^{\rm AOr}({\bf k}) &=&
   \pm t \big[ A_{\bf k} \cos \theta_{-}
  + \gamma \big( C_{\bf k}\cos\theta_{+} + D_{\bf k}\sin\theta_{+} \big)
         \big]          \nonumber \\
   &+&  U\big(\textstyle{\frac{1}{2}}n \pm \hat{G}_{\bf k} \big),
\label{HFevAOreal}
\end{eqnarray}
where
\begin{eqnarray}
\hat{G}_{\bf k} &=& \Big[ T^2 + \Big(\frac{t}{U}\Big)^2
\big\{ A_{\bf k}^2 \sin^2 \theta_{-}                    \nonumber  \\
 && \! + \gamma^2 \big( C_{\bf k} \sin \theta_{+}
                 - D_{\bf k} \cos \theta_{+} \big)^2
      \big\} \Big]^{1/2} ,
\label{Ghat}
\end{eqnarray}
valid in the large $U$ limit ($U/t \gg 1$), and again obtain analytic
self-consistency equations by taking the derivatives of $E^{\rm AOr}$
with respect to $n$, $T$, $\theta_{+}$, and $\theta_{-}$. From the
latter two one can now prove the following. First, that $\theta_{+}=0$,
i.e., $\theta_{\rm A}=-\theta_{\rm B}$, so the pseudospin vectors on the
two sublattices are mirror images of one another with respect to the
cubic direction $\theta=0$ (or the equivalent ones $\theta=\pm 2\pi/3$).
Second, that $\cos \theta_{-} \simeq -(U/3t)x$ for $x\ll t/U$, i.e.,
$\theta_{\rm A}\simeq \pi/2+(U/3t)x$, so that at zero doping the stable
solution is the AO$sa$ state, and with increasing doping the pseudospin
vectors tilt slightly away from the cubic direction, making the
solution gradually resemble more the AO$ab$ state.\cite{noteAOreal}

%%%%%%%%%%%%%%%%%%%%%%%%%%%%%%%%%%%%%%%%%%%%%%%%%%%%%%%%%%%%%%%%%%%%%%%%
%%
%%                             figure 4
%%
%%%%%%%%%%%%%%%%%%%%%%%%%%%%%%%%%%%%%%%%%%%%%%%%%%%%%%%%%%%%%%%%%%%%%%%%
\begin{figure}
\includegraphics[width=7.7cm]{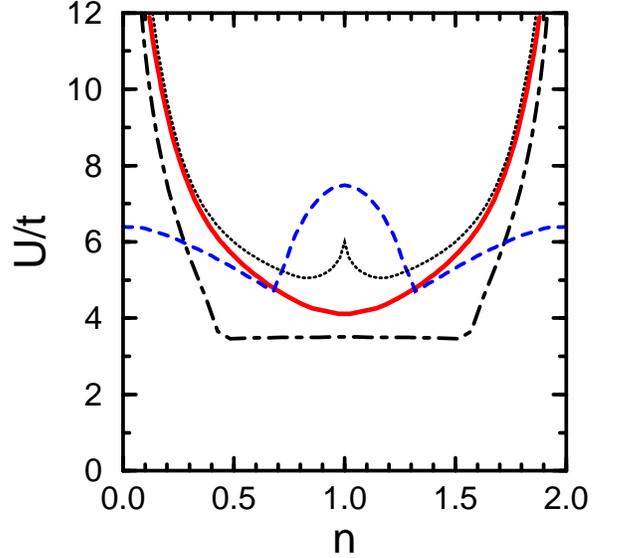}
\caption{(Color online)
Stoner instability towards FO+ partly polarized states (full line) as a
function of band filling $n$ in the orbital Hubbard model ($\gamma=1$),
and the inverse of the $e_g$ density of states of Fig. \ref{fig:dosg}(d)
(dotted line); full polarization occurs only in the limit $U=\infty$.
For the spin model ($\gamma=0$), the corresponding Stoner instability
(dashed-dotted line) is given by the inverse of the density of states
shown in Fig. \ref{fig:dosg}(a), while saturated FM
states occur above the dashed line.
}
\label{fig:stoner}
\end{figure}

As an example of the HF instability at intermediate $U$ we have
investigated how the complex FO+ (or the equivalent FO$-$) state
develops when $U$ increases, using Eqs. (\ref{HamHFFO}) and
(\ref{HFenergyFO}). First, at $\gamma=0$ one recovers the Stoner
criterion $U_oN(E_F)=1$ for the onset of the FM order with increasing
$U$, with the FM saturated states becoming stable at still larger but
finite value of $U$ (Fig. \ref{fig:stoner}). By contrast, in the orbital
model at $\gamma=1$ the instability is qualitatively different, and the
FO+ (FO$-$) state appears as a {\it global property\/} of the band
rather than as an instability at the Fermi surface. The instability
occurs at higher values of $U$ for any filling than in the spin case ---
actually the value of the critical $U$ is very close to that giving full
magnetic polarization in the spin case.

Here, unlike in the spin case, the FO order implies that the electronic
bands are changed --- they develop an additional splitting above a
critical value of $U$, which modifies the shape of the bands and leads
to a finite order parameter $T_z=\langle T_i^z\rangle\neq 0$. This
mechanism of the instability resembles that known in the spin case for
the onset of antiferromagnetism. The critical value of $U$ above which
weak order appears has therefore no relation to the actual shape of the
density of states (see Fig. \ref{fig:stoner}).

We decided not to investigate the phase diagram of the orbital Hubbard
model in the HF approximation in detail. Instead, we concentrate first
on the qualitatively novel aspects of various possible ordered states in
the regime of large $U$, where, as we will see, the contrast with the
spin case manifests itself in the most transparent way. Using these
results, we will then comment of the HF phase diagrams analyzed in
detail by several groups,\cite{Shi00,Bri01,Mae00} in Sec.
\ref{sec:hfphd}.

\subsection{Superexchange in the complex orbital states}
\label{sec:hfco}

We have already seen that the analysis of the orbital-ordered states
simplifies when the splitting of the quasiparticle bands $\propto UT_z$
or $\propto UT$ is sufficiently large that it opens up a gap and only
the lower band (lowest two bands for AO order) is (are) partly occupied
when $n\leq 1$. It is then straightforward to calculate the energy and
the order parameter by summing over the occupied states.

Consider first the ordered states with complex orbitals. In the case of
the FO+ state the equation for the order parameter, from Eqs.~(\ref{THF})
and (\ref{Ehatk}), takes the simple form (because only $T_z\neq 0$, while
$T=0$):
\begin{eqnarray}
\label{HFFOt:op}
T_z&=&\frac{1}{2}\frac{1}{N}\sum_{\bf k}\frac{n_{-}({\bf k})}
        {E_{\bf k}},                                              \\
\label{HFFO:op}
E_{\bf k}&=&\Big[1+\Big(\frac{\gamma t}{UT_z}\Big)^2B_{\bf k}^2\Big]^{1/2}.
\end{eqnarray}
Equation (\ref{HFFO:op}) shows explicitly that, unlike in the spin case,
$T_z=n/2$ only at $U=\infty$, basically because the saturated FO+ state
{\it is not an eigenstate\/} of the orbital Hubbard model given by Eq.
(\ref{H_c}). Thus the FO+ state is again seen to resemble the AF phase
in the spin model.

Similarly, in the AO$\pm$ phase for large enough $U$ the order parameter
is given by
\begin{eqnarray}
\label{HFAOt:op}
T_z&=&\frac{1}{2} \frac{1}{N} \sum_{\bf k}
   \frac{n_{1}({\bf k})+n_{2}({\bf k})}{F_{\bf k}},                 \\
\label{HFAO:op}
F_{\bf k}&=&\Big[1+\Big(\frac{t}{UT_z}\Big)^2 A_{\bf k}^2\Big]^{1/2},
\end{eqnarray}
rather similar to the FO+ case (\ref{HFFO:op}), but with the interchange
$A_{\bf k}\leftrightarrow \gamma B_{\bf k}$. The reason is readily
recognized from Eq.\ (\ref{H_c}): for FO+ order, the diagonal hopping
$\propto c_{i\pm}^{\dagger}c_{j\pm}^{}$ that gives $A_{\bf k}$, is
order-preserving, while the off-diagonal terms
$\propto c_{i\pm}^{\dagger}c_{j\mp}^{}$ that produce $B_{\bf k}$ are
order-perturbing and reduce $T_z$. For AO$\pm$ order this is reversed:
the off-diagonal hopping $\propto\gamma$ that gives $B_{\bf k}$ is
compatible with the order, while the diagonal one that gives
$A_{\bf k}$ disturbs it.

The similarity between the FO+ and AO$\pm$ states at $\gamma\simeq 1$
becomes even more transparent at large $U$ (i.e. $\gg t$), where near
half-filling (i.e. for small $x=1-n>0$), upon expansion up to first
order in $t/U$,
\begin{eqnarray}
\label{HFFOsc:op}
T_z^{{\rm FO}+} &=&   \frac{1}{2}\Big\{(1-x)
   - \frac{3\gamma^2}{(1-x)^2}\Big(\frac{t}{U}\Big)^2\Big\},  \\
\label{HFAOsc:op}
T_z^{{\rm AO}\pm} &=& \frac{1}{2}\Big\{(1-x)
   - \frac{3-2x}{(1-x)^2}\Big(\frac{t}{U}\Big)^2\Big\}.
\end{eqnarray}
Note that a SE contribution $\propto(\gamma t)^2/U$ appears also in the
FO+ state, because the off-diagonal hopping permits virtual charge
fluctuations. This result is again qualitatively different from the spin
case, where the SE contributes only in the AF states, and so
destabilizes uniform FM spin order. In the genuine orbital case
($\gamma=1$) the reduction of the order parameter by SE is the same for
FO+ and AO$\pm$ at $x=0$, but at $x\gtrsim 0$ it is slightly larger for
the FO+ phase. The corresponding expressions for the energy, up to
second order in $t/U$, become
\begin{eqnarray}
& & \hskip -1.2cm
E^{{\rm FO}+} =
    - t \frac{1}{N} \sum_{\bf k} n_{-}({\bf k}) A_{\bf k}
    - \frac{3 \gamma^2}{2(1-x)}\;\frac{t^2}{U},
\label{HFFOsc:energy}   \\
& & \hskip -1.2cm
E^{{\rm AO}\pm} =
    - \gamma t \frac{1}{N} \sum_{\bf k}
     \big[n_{1}({\bf k})-n_{2}({\bf k})\big] B_{\bf k}
        \nonumber \\
& & \hskip 0.2cm
    - \frac{3-2x}{2(1-x)}\;\frac{t^2}{U}.
\label{HFAOsc:energy}
\end{eqnarray}
Both are seen to be composed of the $U=\infty$ kinetic energy (compare
Eqs.~(\ref{dispFO+}) and (\ref{dispAOpm}) for the dispersions) and a
(negative) SE energy. Surprisingly, near half-filling the energy per
site of the FO phase is {\it lower\/} than that of the AO phase at
any value of $U$, not only because the FO phase gains more kinetic
energy $\propto -3tx$ than the AO phase $\propto -2tx$, but also because
it has lower SE energy. Instead, AO$\pm$ order yields lower energy
at larger doping $x\gtrsim 0.27$ as a consequence of its peculiar density
of states [Fig. \ref{fig:dos}(a)].\cite{Shi00,Mae00} Note that this is
{\it opposite\/} to the spin case ($\gamma=0$), where the N\'eel (AF)
state has lower energy near $n=1$ and the FM state takes over only above
a critical doping $x_c\simeq t/2U$.

We emphasize that we have compared as yet only the two complex states
with one another, with the express purpose of contrasting the behavior
of these orbital states with that of the corresponding spin states. To
establish what the most stable orbital-ordered state is, we still have
to consider the real states.

\subsection{Superexchange in the real orbital states}
\label{sec:hfre}

The results obtained for the ordered phases with real orbitals are
qualitatively similar. We focus here on the representative cases of the
FO$x$, the FO$z$, and the ($G$-type) AO$sa$ states, which we have shown
in Section \ref{sec:hf} to be solutions of the HF equations.
Note that the AO$sa$ phase is representative for $G$-type AO order.
For simplicity we ignore here the small higher order correction to the
equations below,\cite{noteAOxpmz} which occur when the actual occupied
orbitals deviate from those of the AO$sa$ state towards those pertaining
to the AO$ab$ state as discussed above.

At large $U/t$ one finds near half-filling for the order parameters
\begin{eqnarray}
& & \hskip -1.4cm
T^{{\rm FO}x(z)} \! = \!
   \frac{1}{2}\Big\{(1-x)
   - \frac{3 \gamma^2}{2(1-x)^2}\Big(\frac{t}{U}\Big)^2\Big\},
\label{HFFOx:op}  \\
& & \hskip -1.2cm
T^{{\rm AO}sa} \! = \!
   \frac{1}{2}\Big\{(1-x) - \frac{6-4x+3\gamma^2}
{2(1-x)^2}\Big(\frac{t}{U}\Big)^2\Big\}.
\label{HFAOxpmz:op}
\end{eqnarray}
The corresponding energies in these ordered phases are

\begin{widetext}
\begin{eqnarray}
E^{{\rm FO}x}  &=&
    - t \frac{1}{N} \sum_{\bf k} n_{-}({\bf k})
     \big(A_{\bf k} - \gamma C_{\bf k}\big)
    - \frac{3 \gamma^2}{4(1-x)}\;\frac{t^2}{U},
\label{HFFOx:energy}   \\
E^{{\rm FO}z}  &=&
    - t \frac{1}{N} \sum_{\bf k} n_{-}({\bf k})
    \big(A_{\bf k} + \gamma C_{\bf k}\big)
    - \frac{3 \gamma^2}{4(1-x)}\;\frac{t^2}{U},
\label{HFFOz:energy}   \\
E^{{\rm AO}sa}  &=&
     - t\frac{1}{N} \sum_{\bf k}
    \big[(n_{1}({\bf k}) - n_{2}({\bf k}) \big] \gamma C_{\bf k}
     - \frac{6-4x+3\gamma^2}{4(1-x)}\;\frac{t^2}{U}.
\label{HFAOxpmz:energy}
\end{eqnarray}
\end{widetext}
Unlike the complex states, the real states are seen not to be degenerate
in the undoped case $x=0$. The AO$sa$ state has the lowest energy
here, even though the SE contributes also in the FO states. However, we
find the same qualitative difference with the familiar AF and FM states
for spin order as we found for the complex orbital states --- again the
SE contributes both in FO and in AO states.

Finally, we remark that the SE contributes also in any other phase,
either with mixed FO and AO order (e.g. in the $C$-AO and $A$-AO phases
of Sec. \ref{sec:oos}), or in a disordered OL state. Depending on
whether the occupied orbitals on a given bond are identical or not,
virtual processes due to pseudospin non-conserving or pseudospin
conserving hopping contribute, and we have verified that qualitatively
similar results are then obtained to those presented in Eqs.
(\ref{HFFOx:op}-\ref{HFAOxpmz:energy}) above. Such terms would play a
role in the low-doping regime and would deserve a separate study in
order to establish the phase diagram of weakly doped manganites. Note
that in that regime also the spin-dependent SE plays a prominent role,
and the present orbital Hubbard model (\ref{H_c}), which
implicitly assumes FM order, becomes insufficient to describe the
physical properties of the real materials. On the other hand, the SE
terms, being all $\propto t^2/U$, vanish in the limit of large $U$
which we consider in Sec. V, and hence they have no consequences for
the stability of the OL phase at $U=\infty$.

\subsection{Qualitative understanding of the Hartree-Fock phase diagram}
\label{sec:hfphd}

Finally, let us analyze the possible instabilities of the orbital
Hubbard model (\ref{H_c}) in the HF approximation. In the large $U$
limit relevant for such instabilities, the total energy consists of the
kinetic energy at $U=\infty$, discussed in Sec. \ref{sec:core}, and a
negative SE energy. While we do not intend to make a quantitative
comparison between the various phases stable in the HF approximation,
knowing that they are anyway destabilized by the correlation effects
(see Sec. \ref{sec:oliq}), this now enables us to get a simple
interpretation of the HF phase diagram of the genuine $e_g$ orbital
model ($\gamma=1$),\cite{Shi00,Bri01,Mae00,She00} using the large $U$
expansion. These earlier HF studies have shown that at half-filling,
and in the regime of small doping, for $U>6t$ the most stable state is
the real `antiferro' orbital state, with the orbitals close to those
found in the AO$sa$ phase. In this regime the SE energy dominates, and
indeed the largest energy gain is then given by Eq.
(\ref{HFAOxpmz:energy}). At increasing hole doping, however, the kinetic
energy of holes moving in the FO$x$ background is much lower than that
in the AO$sa$ phase (see Fig. \ref{fig:Ekin}), leading to a
transition to `ferro' orbital states when the difference between the SE
terms $\propto t^2/U$ is overcome by the difference between the kinetic
energies of these two phases. The region of the AO$sa$ phase in the
phase diagram decreases when the SE gradually looses its importance
with increasing $U$, as shown by the numerical result of Van den Brink
and Khomskii.\cite{Bri01}

At $U=\infty$ the FO$r$ order is found in the HF approximation at any
doping $x>0$. However, at large but finite $U$ the SE is larger in the
FO+ than in either FO$x$ or FO$z$ phase, while the difference in the
kinetic energy is small [Fig. \ref{fig:Ekin}(d)], and thus the FO+ state
is the first stable `ferro' state at intermediate values of $8<U/t<12$
and $x\simeq 0.15$. However, when $x$ increases further, the kinetic
energy difference between the FO$x$ and FO+ phase dominates, and the
orbital order changes to FO$x$. As the SE energy of the two real FO$x$
and FO$z$ states [see Eqs. (\ref{HFFOx:energy}) and (\ref{HFFOz:energy})]
is the same, the difference in the kinetic energy gives a second
transition from the FO$x$ to the FO$z$ phase with increasing $x$.
At small and intermediate $U/t<12$ one finds eventually at $x\sim 0.5$
the AO$\pm$ phase,\cite{Bri01} which is stabilized in this regime by a
combined effect of large SE energy gain and low kinetic energy (see Fig.
\ref{fig:Ekin}) which follows from the peculiar density of states of
this phase.

In a 2D model the phase diagram is quite different,\cite{Tha00} and is
dominated by the generic tendency towards $x^2-y^2$ polarization within
an $(a,b)$ plane.\cite{Mac99} The AO order is then followed by the FO$x$
phase above a critical doping, which decreases with increasing $U/t$. We
note that the region of the FO$x$ phase is enlarged by the offdiagonal
hopping terms $\propto\gamma t$,\cite{Tha00} in agreement with the
above observation that these terms stabilize the FO phases at finite $U$
due to the respective SE energy contributions.

%%%%%%%%%%%%%%%%%%%%%%%%%%%%%%%%%%%%%%%%%%%%%%%%%%%%%%%%%%%%%%%%%%%%%%%%
%%
%%                          Orbital Liquid
%%
%%%%%%%%%%%%%%%%%%%%%%%%%%%%%%%%%%%%%%%%%%%%%%%%%%%%%%%%%%%%%%%%%%%%%%%%
\section{Orbital liquid state}
\label{sec:oliq}

\subsection{Kotliar-Ruckenstein slave boson representation}
\label{sec:krsb}

To understand further the essential differences between orbital and
spin physics, we develop now an approximate description of the
correlated OL disordered state. This is of crucial importance as the
HF approximation permits only a comparison of ordered states with one
another, and therefore does not allow to draw any conclusions concerning
the stability of the orbital-ordered states with respect to disordered
states. This is well known from spin models --- for instance, the FM
states in the 2D Hubbard model are stable only in a narrow range of
doping $x<0.29$ near half-filling,\cite{vdL91} while the HF
approximation predicts FM to be stable at any electron filling $n$.

We will argue below that indeed orbital (FO or AO) order is not robust
at $\gamma=1$ and gets replaced by a disordered (OL) phase, if one goes
{\it beyond\/} the HF approximation and includes electron correlation
effects. As we have already seen, the orbital problem is richer than
the spin case, as various ordered states are nonequivalent when the
SU(2) symmetry is absent. Therefore, we shall consider only the limit of
very strong correlations and investigate the stability of orbital order
specifically in the $U=\infty$ limit, where the OL competes with fully
saturated FO [see Eqs.~(\ref{HFFOsc:op}) and (\ref{HFFOx:op})] and AO
[see Eqs.~(\ref{HFAOsc:op}) and (\ref{HFAOxpmz:op}] states.

In order to obtain a reliable variational method to calculate the
correlation energy, we have followed the slave boson approach introduced
by Kotliar and Ruckenstein \cite{Kot86} for the spin Hubbard model,
and have adapted it to the orbital case.
In this approach the Fock space is enlarged by the introduction of three
auxiliary bosons at each site, one for each local configuration, viz.
$b_{i+}$ and $b_{i-}$ associated with the single-occupancy
configurations $|i+\rangle$ and $|i-\rangle$, and $e_i$ with the empty
configuration $|i0\rangle$ (double occupancy is excluded at $U=\infty$).
Then a physical fermion (electron) $c$ is represented by a pseudofermion
$f$ and two accompanying bosons according to an expression like
$c_{i\beta}^{\dagger} = f_{i\beta'}^{\dagger}b_{i\beta}^{\dagger}e_i$,
where the two bosons keep track of the change of the local configuration
when an electron is added.\cite{Kot86} This construction, however, must
preserve the cubic symmetry of the Hamiltonian (\ref{H_c}), implying
that it has to be gauge invariant with respect to those U(1) rotations
in orbital space that correspond to a permutation of the cubic axes.
The relevant rotation operator is, for arbitrary rotation angle
$\theta$,
\begin{equation}
\hat{U}_i(\theta)=\exp\Big(-i \theta T_i^z \Big).
\label{utheta}
\end{equation}
The complex orbitals pick up just a phase factor under any rotation of
this form, and the operators $\{c_{i+}^{\dagger},c_{i-}^{\dagger}\}$
transform as
\begin{eqnarray}
\hat{U}_i^{}(\theta)c_{i+}^{\dagger}\hat{U}_i^{\dagger}(\theta) &=&
e^{-i\theta/2}c_{i+}^{\dagger},     \nonumber \\
\hat{U}_i^{}(\theta)c_{i-}^{\dagger}\hat{U}_i^{\dagger}(\theta) &=&
e^{+i\theta/2}c_{i-}^{\dagger}.
\label{transc}
\end{eqnarray}
As already indicated in Section \ref{sec:ohm}, the orbital Hubbard
Hamiltonian (\ref{H_c}) is invariant under a uniform rotation at all
sites, if the common rotation angle $\theta$ is one of the three cubic
angles $-4\pi/3$, $+4\pi/3$, $0$, and if this is accompanied by a
corresponding shift of the ``gauge angles'' $\chi_{\alpha}$ by
$+2\pi/3$, $-2\pi/3$, $0$, respectively. Actually the diagonal hopping
terms in (\ref{H_c}) are invariant under the U(1) transformation
(\ref{utheta}) even for arbitrary $\theta$, as a consequence of the
SU(2) symmetry of the spin Hubbard model, while the off-diagonal
hopping terms pick up phase factors,
\begin{eqnarray}
c_{i+}^{\dagger}c_{j-}^{}&\mapsto&e^{-i \theta}
                c_{i+}^{\dagger}c_{j-}^{}, \nonumber \\
c_{i-}^{\dagger}c_{j+}^{}&\mapsto&e^{+i \theta}
                c_{i-}^{\dagger}c_{j+}^{},
\label{rotcc}
\end{eqnarray}
which get compensated by the shift of the $\chi_{\alpha}$ if $\theta$
is a cubic angle. As the three cubic-angle transformations amount to a
forward and to a backward simultaneous cyclic permutation of axes and
orbitals and to the identity, respectively, the invariance expresses the
cubic symmetry of the Hamiltonian.

Therefore, we take the slave boson representation as
\begin{equation}
c_{i\pm}^{\dagger} = b_{i\pm}^{\dagger}f_{i\mp}^{\dagger}e_i^{},
\label{krbosons}
\end{equation}
corresponding to a representation of the local states by
\begin{eqnarray}
|i0\rangle&=&e_i^{\dagger}|{\rm vac}\rangle,   \nonumber \\
|i+\rangle=c_{i+}^{\dagger}|i0\rangle
          &=&b_{i+}^{\dagger}f_{i-}^{\dagger} |{\rm vac}\rangle,
          \nonumber \\
|i-\rangle=c_{i-}^{\dagger}|i0\rangle
          &=&b_{i-}^{\dagger}f_{i+}^{\dagger} |{\rm vac}\rangle,
\label{krstates}
\end{eqnarray}
and we {\it impose}\/ that the boson and pseudofermion operators
transform under U(1) rotations \cite{noteU1} as
\begin{eqnarray}
\hat{U}_i(\theta) \: e_i^{\dagger} \: \hat{U}_i^{\dagger}(\theta)
&=& e_i^{\dagger},                          \nonumber \\
\hat{U}_i(\theta)b_{i\pm}^{\dagger}\hat{U}_i^{\dagger}(\theta) &=&
e^{\mp i\theta} \,  b_{i\pm}^{\dagger}, \nonumber \\
\hat{U}_i(\theta)f_{i\pm}^{\dagger}\hat{U}_i^{\dagger}(\theta) &=&
e^{\mp i\theta/2}f_{i\pm}^{\dagger}.
\label{transb}
\end{eqnarray}
Note that the phase of the boson operators $b_{i\pm}^{\dagger}$
changes twice as fast as the phase of the pseudofermion operators
$f_{i\pm}^{\dagger}$, i.e. the bosons have pseudospin $T=1$, while the
(pseudo)fermions belong to $T=1/2$. This property guarantees that the
U(1) rotation behavior of the electron operators, as given in
Eqs. (\ref{transc}), is correctly reproduced by the transformation
(\ref{krbosons}). Thus the present formulation is indeed gauge invariant
and preserves the cubic symmetry of the orbital problem, like the
SU(2)-invariant formulation introduced by Fr\'{e}sard and W\"{o}lfle
preserves the full rotational symmetry for the spin system.\cite{Fre92}
Clearly, the construction of a gauge invariant formulation is greatly
facilitated by our use of the complex-orbital representation, but a
similarly gauge invariant representation in terms of real operators can
also be constructed, and is given in the Appendix.

The enlarged Fock space contains also unphysical states which must be
eliminated by imposing constraints as in the original formulation by
Kotliar and Ruckenstein,\cite{Kot86}
\begin{eqnarray}
b_{i+}^{\dagger}b_{i+}^{}+b_{i-}^{\dagger}b_{i-}^{}
+e_{i}^{\dagger}e_{i}^{}&=&1,                          \nonumber \\
b_{i+}^{\dagger}b_{i+}^{} = f_{i-}^{\dagger}f_{i-}^{}, \hskip 0.5cm
b_{i-}^{\dagger}b_{i-}^{}&=&f_{i+}^{\dagger}f_{i+}^{},
\label{const}
\end{eqnarray}
and implemented by means of Lagrange multiplyers $\{\lambda_i, \mu_{i+},
\mu_{i-}\}$. The first constraint excludes double occupancy, the other
two eliminate the unphysical singly-occupied states
$b_{i+}^{\dagger}f_{i+}^{\dagger}|{\rm vac}\rangle$ and
$b_{i-}^{\dagger}f_{i-}^{\dagger}|{\rm vac}\rangle$.
The electron density and the $z$-component of the pseudospin can then be
described at each site either by slave boson or by pseudofermion
operators,
\begin{eqnarray}
& & \hskip -1.1cm
n_i\! \equiv c_{i+}^{\dagger}c_{i+}^{}+c_{i-}^{\dagger}c_{i-}^{}
                        \nonumber \\
& & \hskip -0.72cm
   = b_{i+}^{\dagger}b_{i+}^{}+b_{i-}^{\dagger}b_{i-}^{}
   = f_{i+}^{\dagger}f_{i+}^{}+f_{i-}^{\dagger}f_{i-}^{} ,
\label{density}       \\
& & \hskip -1.2cm
T_i^z\! = \textstyle{\frac{1}{2}}
   (b_{i+}^{\dagger}b_{i+}^{}\!- b_{i-}^{\dagger}b_{i-}^{})
 = \textstyle{\frac{1}{2}}
   (f_{i-}^{\dagger}f_{i-}^{}\!- f_{i+}^{\dagger}f_{i+}^{}).
\label{pseudo-z}
\end{eqnarray}
The other two components of the pseudospin operator can only be
represented as
\begin{eqnarray}
T_i^{+}= b_{i+}^{\dagger}b_{i-}^{} f_{i-}^{\dagger}f_{i+}^{} ,
\label{pseudo+}                \\
T_i^{-}= b_{i-}^{\dagger}b_{i+}^{} f_{i+}^{\dagger}f_{i-}^{} ,
\label{pseudo-}
\end{eqnarray}
and cannot be reduced to expressions in terms of either slave bosons or
pseudofermions alone.\cite{note-T+T-}

As in the spin case one further has to renormalize the bosonic factor
in Eq.~(\ref{krbosons}) in order to recover, when a mean-field
approximation is going to be made and the constraints are no longer
rigorously obeyed, the correct unrenormalized hopping for the
pseudofermions in the uncorrelated ($U=0$) limit. The renormalized
boson factors take the form
\begin{equation}
z_{i\pm}^{\dagger}=\frac{b_{i\pm}^{\dagger}e_i^{}}
{\sqrt{(1-e_i^{\dagger}e_i^{}-b_{i\mp}^{\dagger}b_{i\mp}^{})
      (1-b_{i\pm}^{\dagger}b_{i\pm}^{})}},
\label{zi}
\end{equation}
where it is important that the operator expression under the square root
in the denominator is U(1) invariant, so that $z_{i+}^{\dagger}$
($z_{i-}^{\dagger}$) transforms under (\ref{transb}) exactly as
$b_{i+}^{\dagger}$ ($b_{i-}^{\dagger}$). Then the Hamiltonian in the
slave boson representation at $U=\infty$ becomes
\begin{eqnarray}
{\cal{H}}_{U=\infty}&=&-\frac{1}{2}t
 \sum_{\alpha} \sum_{\langle ij\rangle \parallel \alpha}\Big[
 z_{i+}^{\dagger}f_{i-}^{\dagger}f_{j-}^{}z_{j+}^{}
 +z_{i-}^{\dagger}f_{i+}^{\dagger}f_{j+}^{}z_{j-}^{}
              \nonumber \\
&+& \!\! \gamma \Big( e^{-i\chi_{\alpha}}
 z_{i+}^{\dagger}f_{i-}^{\dagger}f_{j+}^{}z_{j-}^{}
            + e^{+i\chi_{\alpha}}
 z_{i-}^{\dagger}f_{i+}^{\dagger}f_{j-}^{}z_{j+}^{}\Big)\Big]
              \nonumber \\
  &-& \sum_{i}\lambda_{i} \Big(
b_{i+}^{\dagger}b_{i+}^{}+b_{i-}^{\dagger}b_{i-}^{}
+e_{i}^{\dagger}e_{i}^{}-1 \Big)
              \nonumber \\
&-& \mu \sum_{i\lambda} f_{i\lambda}^{\dagger}f_{i\lambda}^{}
   +\sum_{i\lambda} \mu_{i \lambda} \Big(
   b_{i\lambda}^{\dagger}b_{i\lambda}^{}
     - f_{i \bar{\lambda}}^{\dagger}f_{i \bar{\lambda}}^{} \Big) ,
\label{hkr}
\end{eqnarray}
with $\lambda=\pm$ and $\bar{\lambda}=-\lambda$.
The Hamiltonian commutes with the constraints and thus does not connect
the physical and the unphysical subspaces of Fock space.

In the mean-field approximation we replace the boson operators by
their averages. In order not to spoil the cubic invariance only their
amplitudes are replaced by c-numbers, while their phases are prescribed
to behave still according to Eq.~(\ref{transb}).\cite{notepath} So we
set for the boson invariants
\begin{eqnarray}
\langle b_{i+}^{\dagger} b_{i+}^{} \rangle &\equiv& \bar{b}_{i+}^2,
\nonumber \\
\langle b_{i-}^{\dagger} b_{i-}^{} \rangle &\equiv& \bar{b}_{i-}^2,
\nonumber \\
\langle e_{i}^{\dagger} e_{i}^{} \rangle &\equiv& \bar{e}_{i}^2,
\label{bbar}
\end{eqnarray}
where $\bar{b}_{i+}$, $\bar{b}_{i-}$, and $\bar{e}_{i}$ are real
quantities, i.e. do not contain any nontrivial phase.\cite{Fre92}
For the offdiagonal, noninvariant, two-boson products we set
\begin{eqnarray}
& & \hskip -1.2cm
\langle b_{i+}^{\dagger} e_{i}^{} \rangle \equiv
     \bar{b}_{i+}\bar{e}_i \; e^{-i\hat{\vartheta}_i} , \: \:
\langle e_{i}^{\dagger} b_{i+}^{} \rangle \equiv
     \bar{b}_{i+}\bar{e}_i \; e^{+i\hat{\vartheta}_i} ,
                        \nonumber \\
& & \hskip -1.2cm
\langle b_{i-}^{\dagger} e_{i}^{} \rangle \equiv
     \bar{b}_{i-}\bar{e}_i \; e^{+i\hat{\vartheta}_i} , \: \:
\langle e_{i}^{\dagger} b_{i-}^{} \rangle \equiv
     \bar{b}_{i-}\bar{e}_i \; e^{-i\hat{\vartheta}_i} ,
\label{bbaroffd}
\end{eqnarray}
where the `phase operator' $\hat{\vartheta}_i$ is understood
to transform as
\begin{equation}
\hat{U}_i(\theta) \; \hat{\vartheta}_i \;
\hat{U}_i^{\dagger}(\theta) = \hat{\vartheta}_i + \theta,
\label{tranvartheta}
\end{equation}
and in particular assumes the cubic values $\vartheta_a$,
$\vartheta_b$, and $\vartheta_c$ when the two-boson operator product
occurs in an expression taken along the $a$-axis, $b$-axis, or
$c$-axis, respectively. The last average of Eqs. (\ref{bbar})
controls the number of holes in the $e_g$ band,
$\bar{e}_{i}^2=x$, for a phase with uniform charge density. The
constraints give then the following self-consistency conditions,
\begin{eqnarray}
\bar{b}_{i+}^2=\langle f_{i-}^{\dagger}f_{i-}^{} \rangle,
\hskip 0.5cm  &&
\bar{b}_{i-}^2 = \langle f_{i+}^{\dagger}f_{i+}^{} \rangle,
\nonumber \\
\bar{b}_{i+}^2+\bar{b}_{i-}^2 &=& 1-x,
\label{bbarx}
\end{eqnarray}
while the renormalization factors become
\begin{equation}
\langle z_{i\pm}^{\dagger} \rangle \equiv \sqrt{q_{i\pm}} \;
   e^{\mp i \hat{\vartheta}_i},  \;\;
\langle z_{i\pm}^{}\rangle   \equiv \sqrt{q_{i\pm}} \;
   e^{\pm i \hat{\vartheta}_i} ,
\label{renormMF}
\end{equation}
with
\begin{equation}
q_{i\pm}=\frac{x}{1-\langle f_{i\mp}^{\dagger}f_{i\mp}^{} \rangle}
   =\frac{x}{1-\langle n_{i\pm} \rangle}.
\label{gutzwiller}
\end{equation}
The exponentials containing $\hat{\vartheta}_i$ can be eliminated from
the Hamiltonian by absorbing them in the pseudo-fermions, according to
\begin{equation}
\hat{f}_{i\pm}^{\dagger} =
       e^{\mp i \hat{\vartheta}_i} f_{i\mp}^{\dagger}.
\label{barf}
\end{equation}
Note that this definition ensures that the $\hat{f}_{i\pm}^{\dagger}$
transform properly under U(1) in accordance with Eq.~(\ref{transb}).

Within the slave boson mean-field approximation one thus finds an
effective Hamiltonian for pseudofermions subject to local constraints,
and with renormalized hopping. In the case of orbital-ordered phases
its precise form depends on the assumed type of state, with the hopping
renormalization factors $q_{i\pm}^{}$ either uniform or alternating
between two sublattices. Here we present only its simpler form,
adequate for uniform phases, such as FO and OL states, in which the
renormalization factors and Lagrange parameters can be taken
site independent,
\begin{eqnarray}
{\cal H}_{U=\infty}^{\rm MF}&=&
-\frac{1}{2}t\sum_{\alpha}\sum_{\langle ij\rangle\parallel\alpha}
   \Big[q_+^{}\hat{f}_{i+}^{\dagger}\hat{f}_{j+}^{}
   +q_-^{}\hat{f}_{i-}^{\dagger}\hat{f}_{j-}^{}
                                                   \nonumber \\
&+&\gamma\sqrt{q_+q_-}\Big(e^{-i\chi_{\alpha}}
 \hat{f}_{i+}^{\dagger}\hat{f}_{j-}^{}
  +e^{+i\chi_{\alpha}}\hat{f}_{i-}^{\dagger}\hat{f}_{j+}^{}\Big)\Big]
                                                   \nonumber \\
&-&\sum_{i\lambda}\mu_{\lambda} \hat{n}_{i\lambda},
\label{hkrmf}
\end{eqnarray}
with $\hat{n}_{i\lambda} =
 \hat{f}_{i\lambda}^{\dagger}\hat{f}_{i\lambda}^{}$.
The present formalism reproduces the results of Kotliar and
Ruckenstein for the spin model ($\gamma=0$) with hopping $\frac{1}{2}t$,
and gives the same results as the Gutzwiller approximation,\cite{Gut65}
and so $q_{+}$ and $q_{-}$ will be called also Gutzwiller factors.

The ordered states can be obtained within the present KR slave boson
approach by a proper choice of the Lagrange multipliers. For instance,
the FO+ state is now obtained from Eq. (\ref{hkrmf}) by imposing
$\langle \hat{n}_{i-}\rangle=0$ by means of the condition
$\mu_{-}=-\infty$
(while $\mu_{+}=0$). Such states do not experience any band narrowing,
as double occupancy is rigorously eliminated at $U=\infty$, and the
correlation energy vanishes.\cite{note:slavefermion} As a result, only
the $\varepsilon_{U=\infty}^{\rm FO}({\bf k})=-tA_{\bf k}$ band
is partly filled in the FO+ state, while the
$\varepsilon_{U=\infty,\pm}^{\rm AO}({\bf k})=\pm\gamma tB_{\bf k}$
bands are filled in the AO$\pm$ state. Real orbital-ordered states can
also be obtained, using the formalism described in the Appendix.
Therefore, in the $U=\infty$ limit one reproduces the results of the
HF approximation described for these states in Sec.~\ref{sec:hf}.

\subsection{Nature of the orbital liquid state}
\label{sec:nature}

%%%%%%%%%%%%%%%%%%%%%%%%%%%%%%%%%%%%%%%%%%%%%%%%%%%%%%%%%%%%%%%%%%%%%%%%
%%
%%                             figure 5
%%
%%%%%%%%%%%%%%%%%%%%%%%%%%%%%%%%%%%%%%%%%%%%%%%%%%%%%%%%%%%%%%%%%%%%%%%%
\begin{figure}
\includegraphics[width=7.7cm]{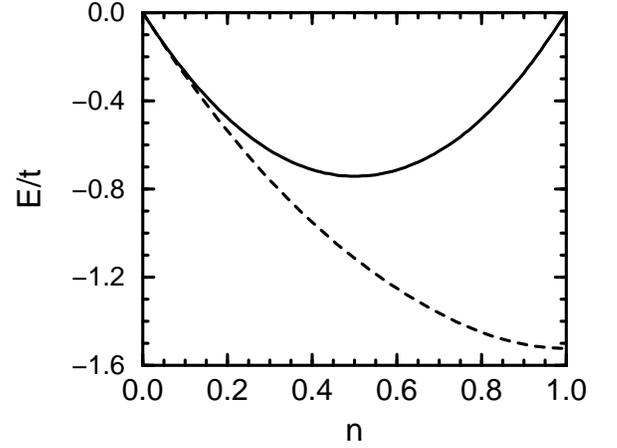}
\caption{
Kinetic energies $E/t$ of the OL state for uncorrelated
($U=0$, dashed line) and correlated ($U=\infty$, full line) $e_g$
electrons (at $\gamma=1$), as functions of the electron density $n$.
}
\label{fig:oliq}
\end{figure}
%%%%%%%%%%%%%%%%%%%%%%%%%%%%%%%%%%%%%%%%%%%%%%%%%%%%%%%%%%%%%%%%%%%%%%%%

A qualitatively new solution, however, is obtained within the present
approximation for the {\it disordered\/} state, where double occupancies
are on average eliminated by the slave bosons, and this correlation
effect leads to an increase of the kinetic energy. The minimum energy
is obtained when the pseudofermion densities are equal,
$\langle \hat{n}_{i+}\rangle
=\langle \hat{n}_{i-}\rangle=\frac{1}{2}(1-x)$,
and the Gutzwiller renormalization factors take the simple form,
\begin{equation}
q(x)=q_{\pm}(x)=\frac{2x}{1+x}.
\label{qreno}
\end{equation}
Then the pseudofermion bands,
\begin{eqnarray}
\varepsilon_{U=\infty,\pm}^{\rm OL}({\bf k}) &=&
    q(x) \: \varepsilon_{U=0,\pm}({\bf k})
                \nonumber \\
&=& -tq(x)\big[A_{\bf k}\pm\gamma B_{\bf k}\big],
\label{feba}
\end{eqnarray}
represent formally the superposition of the FO+ and AO$\pm$ bands given
by Eq. (\ref{disp_superpos}), typical for uncorrelated $e_g$ electrons,
but now {\it renormalized by correlations\/}. They interpolate
correctly between the case of uncorrelated electrons in an empty band
($x\simeq 1$) and a Mott insulator at half-filling ($x=0$) where the
dispersion is fully suppressed, as illustrated in Fig. \ref{fig:oliq}.
Owing to the Gutzwiller factors the kinetic energy has a minimum at
filling $n=0.5$, and approaches zero at $n=1$. Thus, the kinetic energy
has a similar doping dependence to that found in a spinless
fermion model, i.e., for fermions with a {\it single\/} orbital flavor.
As in the spin case,\cite{Joz88} one can argue that at $x\sim 0$ strong
correlations lead to an effective exclusion principle between the two
degrees of freedom also in ${\bf k}$ space, i.e., for each momentum
${\bf k}$ only one orbital flavor may be occupied.

This OL state is fully isotropic in the sense that the mean-field values
of the pseudospin operators vanish, i.e.,
\begin{equation}
\langle T_i^x \rangle = \langle T_i^y \rangle =
   \langle T_i^z \rangle = 0 .
\label{Tavzero}
\end{equation}
For the $z$-component this follows immediately from Eq.~(\ref{pseudo-z})
once $\bar{b}_{i+}^2=\bar{b}_{i-}^2$. For the other components we apply
Eqs. (\ref{bbaroffd}) and (\ref{barf}) to Eq.~(\ref{pseudo+}) and obtain
\begin{eqnarray}
\langle T_i^{+} \rangle &=&
\bar{b}_{i+} \bar{b}_{i-} \; e^{-2i\hat{\vartheta}_i}
       \langle f_{i-}^{\dagger} f_{i+}^{} \rangle
                \nonumber \\
        &=& \textstyle{\frac{1}{2}} (1-x)
       \langle \hat{f}_{i+}^{\dagger} \hat{f}_{i-}^{} \rangle ,
\label{pseudo+av}
\end{eqnarray}
and similarly for $\langle T_i^{-} \rangle$. The pseudofermion averages
can be determined by making use of Fourier transformation:
since the Fourier-transformed Hamiltonian (\ref{hkrmf}) can be
diagonalized analytically, the Fourier-transformed pseudofermion
operators can be expressed in terms of the eigenvectors
$\{ e_{{\bf k}+}, e_{{\bf k}-} \}$, with the result
\begin{eqnarray}
\langle \hat{f}_{{\bf k}+}^{\dagger} \hat{f}_{{\bf k}-}^{} \rangle\! +\!
\langle \hat{f}_{{\bf k}-}^{\dagger} \hat{f}_{{\bf k}+}^{} \rangle
\!&=&\! \frac{C_{\bf k}}{B_{\bf k}}
\Big( \langle e_{{\bf k}+}^{\dagger} e_{{\bf k}+}^{} \rangle
    \!-\! \langle e_{{\bf k}-}^{\dagger} e_{{\bf k}-}^{} \rangle \Big) ,
                \nonumber \\
\langle \hat{f}_{{\bf k}+}^{\dagger} \hat{f}_{{\bf k}-}^{} \rangle\! -\!
\langle \hat{f}_{{\bf k}-}^{\dagger} \hat{f}_{{\bf k}+}^{} \rangle
\!&=&\! i \frac{D_{\bf k}}{B_{\bf k}}
\Big( \langle e_{{\bf k}+}^{\dagger} e_{{\bf k}+}^{} \rangle
    \!+\! \langle e_{{\bf k}-}^{\dagger} e_{{\bf k}-}^{} \rangle \Big) .
                \nonumber \\
& &
\label{barfkav}
\end{eqnarray}
Since the eigenvalues $\varepsilon_{U=\infty,\pm}^{\rm OL}({\bf k})$
are cubic invariant [see Eq.~({\ref{feba})] in each of the
two bands the three states with the components of ${\bf k}$ cyclically
permuted are either all occupied or all unoccupied, and thus
\begin{equation}
C_{+} = \sum_{\bf k} \frac{ \cos k_{\alpha} }{B_{\bf k} }
   \langle e_{{\bf k}+}^{\dagger} e_{{\bf k}+}^{} \rangle
\label{sumplus}
\end{equation}
is independent of $\alpha$, and similarly for $C_{-}$. It then follows
from the form of $C_{\bf k}$ and $D_{\bf k}$ [see Eqs.~(\ref{dispCk})
and (\ref{dispDk})] that the expressions
$\langle\hat{f}_{{\bf k}\pm}^{\dagger}\hat{f}_{{\bf k}\mp}^{}\rangle$,
given by Eqs. (\ref{barfkav}), both give zero
when summed over the Brillouin zone, and so
\begin{equation}
\langle \hat{f}_{i+}^{\dagger} \hat{f}_{i-}^{} \rangle =
 \langle \hat{f}_{i-}^{\dagger} \hat{f}_{i+}^{} \rangle = 0 ,
\label{barfavzero}
\end{equation}
and Eq.~(\ref{Tavzero}) follows.

The absence of a preferred orientation of the pseudospin implies that
there is no orbital preferentially occupied. In particular,
$\langle T_i^x \rangle = 0$ and $\langle T_i^y \rangle = 0$ imply
[see Eq.~(\ref{complex})] that
\begin{eqnarray}
\langle c_{iz}^{\dagger} c_{iz}^{}
   - c_{ix}^{\dagger} c_{ix}^{} \rangle &=& 0 ,
        \nonumber \\
\langle c_{iz}^{\dagger} c_{ix}^{}
   + c_{ix}^{\dagger} c_{iz}^{} \rangle &=& 0 ,
\label{czcxzero}
\end{eqnarray}
from which it follows that the same relations hold for the operators
$\{{c'}_{iz}^{'\dagger},{c'}_{ix}^{\dagger}\}$ obtained after an
arbitrary U(1) rotation, as is easily verified explicitly or by
observing that $T_i^x$ and $T_i^y$ rotate as an $E$ doublet [compare
the Appendix]. Thus, the OL is SU(2) symmetric --- random complex or
random real orbitals are equivalent, and indeed the {\it identical\/}
OL state is obtained using real orbitals, as shown in the Appendix.
This {\it correlated disordered\/} OL state with completely randomly
occupied orbitals is apparently different from that proposed by
Ishihara, Yamanaka, and Nagaosa,\cite{Nag97} in which the planar
orbitals $\{x^2-y^2$, $y^2-z^2$, $z^2-x^2\}$ play a prominent role.

\subsection{Absence of the Nagaoka theorem}
\label{sec:nagaoka}

Before investigating the stability of the OL state in Sec.
\ref{sec:staliq}, let us consider the special case of a single hole in
a half-filled system. In the spin case ($\gamma=0$) the celebrated
Nagaoka theorem,\cite{Nag66} one of the very few exact results in the
theory of itinerant magnetism, then applies: Nagaoka has shown that the
ground state is FM when a single hole/electron is added to a
half-filled system, described by the spin Hubbard model at $U=\infty$.
A central assumption of this theorem is that the kinetic energy
conserves the spin flavor (see, e.g., the proof in Ref.
\onlinecite{Nag66}), precisely the feature not obeyed by the orbital
flavor of $e_g$ electrons. Thus, at $\gamma\neq 0$ no exact statement
can be made for the orbital Hubbard model (\ref{H_c}) and,
{\it a priori\/}, one expects that polarized states are harder to
stabilize in this case.

%%%%%%%%%%%%%%%%%%%%%%%%%%%%%%%%%%%%%%%%%%%%%%%%%%%%%%%%%%%%%%%%%%%%%%%%
%%
%%                             figure 6
%%
%%%%%%%%%%%%%%%%%%%%%%%%%%%%%%%%%%%%%%%%%%%%%%%%%%%%%%%%%%%%%%%%%%%%%%%%
\begin{figure}
\includegraphics[width=8.2cm]{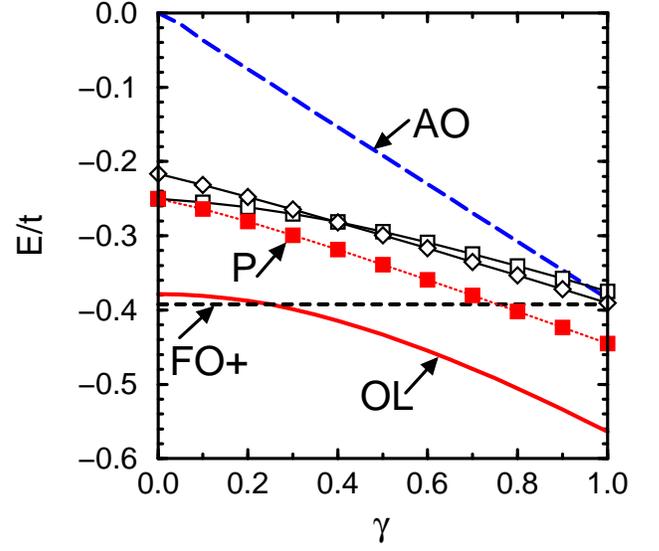}
\caption{(Color online)
Kinetic energies $E$ per site at electron density $n=0.75$ and
$U=\infty$ for increasing
off-diagonal hopping $\propto\gamma$ in Eq. (\protect\ref{H_c}), as
obtained in the KR approach for: the OL ground state (solid line),
FO+ (dashed line), AO$\pm$ state (AO, long-dashed line), and
energy $E_{\Box}$ (filled squares) for the ground state of a four-site
plaquette (P).
Also shown are the energies of the lowest two excited states for the
plaquette: a nondegenerate state which splits off the degenerate ground
state at $\gamma=0$ (empty squares), and a doubly degenerate state with
finite excitation energy at $\gamma=0$ (diamonds).
}
\label{fig:ekin}
\end{figure}

We have investigated the consequences of the SU(2) symmetry breaking,
i.e. of the pseudospin non-conservation, by analyzing the exact solution
for a plaquette (four-site cluster) filled by three electrons,
as a function of $\gamma$. In the spin model, at $\gamma=0$, the ground
state, with kinetic energy $E_{\Box}=-0.25t$ per site, is fourfold
degenerate, corresponding to maximum spin $S=\frac{3}{2}$ as required by
the Nagaoka theorem. At $\gamma>0$ it splits into four nondegenerate
states: the ground state and three excited states (the lowest of them is
shown in Fig. \ref{fig:ekin}). The first excited state in the spin model
($\gamma=0$) is doubly degenerate, and this degeneracy is not removed at
$\gamma>0$, and the two states lower their energy when $\gamma$
increases towards $\gamma=1$. For $\gamma\ge 0.4$ this degenerate
excited state has already a lower energy than any other excited state
(the level crossing is shown in Fig. \ref{fig:ekin}). None of these
states can be classified by a pseudospin quantum number. In the genuine
orbital case ($\gamma=1$) the kinetic energy per site in the ground
state, $E_{\Box}\simeq -0.44t$, is much lower than in the spin case
(at $\gamma=0$), showing that a considerable amount of kinetic energy is
gained when the orbitals get disordered and full advantage is taken of
the pseudospin non-conserving hopping. This result suggests that a
similar tendency towards disorder should be present in the thermodynamic
limit.

\subsection{Stability of the orbital liquid phase}
\label{sec:staliq}

Also for the full 3D model it is instructive to consider, at fixed
density, the variation with $\gamma$ of the total energy $E$ of possible
ordered and disordered states. We do so in Fig.~\ref{fig:ekin} at
the same filling $n=0.75$ as one has in the plaquette filled by three
electrons, in order to enable a comparison with the exact results for
that finite system. The energy of the polarized FO+ state does not
depend on $\gamma$ [see Eq.~(\ref{dispFO+})], while that of the AO$\pm$
state follows from the dispersion given by Eq. (\ref{dispAOpm}), and
decreases linearly with $\gamma$. At $\gamma=1$ it comes very close to
that of the FO+ state, but remains still a little bit higher.
At $\gamma=0$ the polarized FO+ phase has a lower energy than the OL
state, which confirms that FM states are stable in a range of filling
close to $n=1$ in the 3D Hubbard model.\cite{Moe93} The energy of the
OL phase decreases gradually with increasing $\gamma$, and becomes
lower than that of the FO+ phase (which stays constant) at
$\gamma\simeq 0.25$. It is remarkable that the energy decrease in the
OL phase, when going from $\gamma=0$ to $\gamma=1$, is quite large,
and actually of similar magnitude as the exact result in the finite
system. Hence, one finds that in spite of the renormalization of the
hopping by $q(1/4)=0.4$, the (kinetic) energy in the OL state is
substantially lower than in the AO$\pm$ state.

%%%%%%%%%%%%%%%%%%%%%%%%%%%%%%%%%%%%%%%%%%%%%%%%%%%%%%%%%%%%%%%%%%%%%%%%
%%
%%                             figure 7
%%
%%%%%%%%%%%%%%%%%%%%%%%%%%%%%%%%%%%%%%%%%%%%%%%%%%%%%%%%%%%%%%%%%%%%%%%%
\begin{figure}
\includegraphics[width=8.2cm]{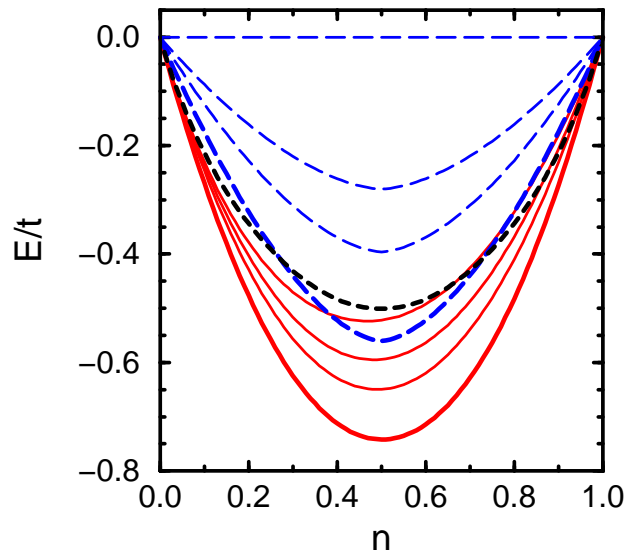}
\caption{(Color online)
Kinetic energy $E$ in the KR mean-field approximation as
functions of $n$ for: AO$\pm$ (long-dashed lines) and OL state (full
lines) for increasing $\gamma=0$, 0.5, 0.707, and 1 from top to bottom;
the dashed line shows the kinetic energy of the FO+ state which is
independent of $\gamma$.
}
\label{fig:kr}
\end{figure}

Next we consider the variation of the total energy $E$ of ordered and
disordered {\it complex\/} orbital states with electron filling $n$
(Fig. \ref{fig:kr}). In the spin model ($\gamma=0$) the FM phase has
somewhat lower energy than the disordered OL state close to half-filling,
in the range $n>2/3$.\cite{notekr} Our approach reproduces in this limit
the known result of the slave boson approach, which gives a FM ground
state for any bipartite lattice with the density of states being an even
function of energy.\cite{Moe93} When $\gamma$ is increased,
$E_{\rm FO}$ does not change, whereas $E_{{\rm AO}\pm}$, initially at
zero for $\gamma=0$, decreases $\propto\gamma$, and at $\gamma=1$
surpasses the FO+ state at $x=0.27$. Hence, the slave boson approach
reproduces here the result of the HF approximation for these states.
\cite{Tak98} However, in spite of the band narrowing $\propto q(x)$,
which is appreciable at these electron densities near half-filling,
considerably {\it more (kinetic) energy is gained in the OL state\/}.
This is basically due to the fact that both hopping channels
contribute, which gives rise to the large density of states over the
full frequency range, and at small doping in particular [compare
Fig.~\ref{fig:dosg}(d) with Fig.~\ref{fig:dos}]. We may conclude that
the presence of the additional non-pseudospin-conserving hopping
channel, associated with the absence of SU(2) symmetry, implies that
more kinetic energy can be gained by paying correlation energy than
in the spin case, and that this favors the disordered OL state
sufficiently to make its energy lower than those of the complex
orbital-ordered states at any value of $n$.

%%%%%%%%%%%%%%%%%%%%%%%%%%%%%%%%%%%%%%%%%%%%%%%%%%%%%%%%%%%%%%%%%%%%%%%%
%%
%%                             figure 8
%%
%%%%%%%%%%%%%%%%%%%%%%%%%%%%%%%%%%%%%%%%%%%%%%%%%%%%%%%%%%%%%%%%%%%%%%%%
\begin{figure}
\includegraphics[width=7.7cm]{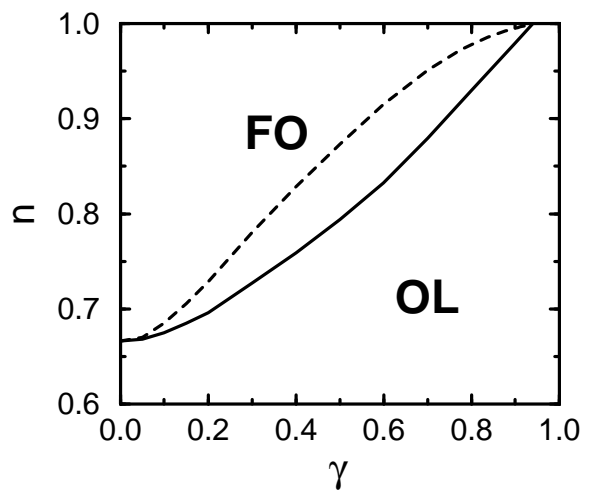}
\caption{
Region of stability of the FO states at $U=\infty$ as a function of
$\gamma$; the transition to the OL state from the FO$x(z)$ and complex
FO+ state are shown by the full and dashed line, respectively.
}
\label{fig:phd}
\end{figure}

Finally we compare at $U=\infty$ the energies of {\it all\/} states,
both with complex and real orbitals, varying $n$ and $\gamma$. One finds
that AO states are never stable in this limit of strong correlation,
while FO states are stable only at small $\gamma$ [Fig. \ref{fig:phd}].
At $\gamma=0$ (the spin case) the FO+ and FO$x$ (FO$z$) states are
necessarily degenerate, but at any $\gamma>0$ the phases with ordered
real orbitals have lower energy, with FO$z$ (FO$x$) being more stable at
$n<0.71$ ($n>0.71$). The range of FO order shrinks gradually with
increasing $\gamma$, and {\it above $\gamma\simeq 0.94$ the OL phase is
stable in the entire range of $n$\/}. We argue that at finite $U$ the
kinetic energy will become even more dominant and thus will strongly
favor disorder, except near $n\simeq 1$ where SE stabilizes real-orbital
AO order.\cite{Kug82,Ole03,Fei97,Fei99,Mae00} We thus conclude that for
the $e_g$ orbital Hubbard model ($\gamma=1$) doping triggers a crossover
to the OL state {\it at any $U$\/}, supporting earlier conjectures that
such a disordered state is realized.\cite{Nag97,Kil98}

\subsection{Brinkman-Rice transition at $n=1$}
\label{sec:mit}

At half-filling ($n=1$) it is straightforward to apply the finite-$U$
version of the KR formalism,\cite{Kot86} and investigate the generic
metal-insulator transition in the orbital disordered phase, ignoring
the AO order promoted by the SE. Here one introduces as a counterpart
to the bosons $e_i$ which control the empty configurations
$|0\rangle=e_i^{\dagger}|{\rm vac}\rangle$, also bosons $d_i$ which
control the double occupancies
$c_{i \uparrow}^{\dagger}c_{i \downarrow}^{\dagger}|0\rangle=
d_i^{\dagger}f_{i \uparrow}^{\dagger}f_{i \downarrow}^{\dagger}
|{\rm vac}\rangle$. The mean-field approximation gives then the
renormalization factor (at $n=1$),\cite{Kot86}
\begin{equation}
\eta(d)=8d^2(1-2d^2),
\label{qbr}
\end{equation}
where $d=\langle d_i \rangle$ is the average amplitude of a doubly
occupied configuration in the ground state. The bands are then given
by the dispersion for free electrons (\ref{disp_unc}) renormalized by
$\eta(d)$,
\begin{equation}
\varepsilon_{U,\pm}^{\rm OL}({\bf k})
= \eta(d) \varepsilon_{U=0,\pm}({\rm k})
=-\eta(d)t [ A_{\bf k}\pm\gamma B_{\bf k} ].
\label{disp_olu}
\end{equation}
So the kinetic energy is $\eta(d)\bar{\epsilon}_0(\gamma)$, where
$\bar{\epsilon}_0(\gamma)$ is the kinetic energy of the uncorrelated
OL, obtained by integrating the two bands
$\varepsilon_{U=0,\pm}({\rm k})$ (\ref{disp_unc}) up to half-filling,
while the Coulomb repulsion gives an energy $Ud^2$ per site.

%%%%%%%%%%%%%%%%%%%%%%%%%%%%%%%%%%%%%%%%%%%%%%%%%%%%%%%%%%%%%%%%%%%%%%%%
%%
%%                             figure 9
%%
%%%%%%%%%%%%%%%%%%%%%%%%%%%%%%%%%%%%%%%%%%%%%%%%%%%%%%%%%%%%%%%%%%%%%%%%
\begin{figure}
\includegraphics[width=7.7cm]{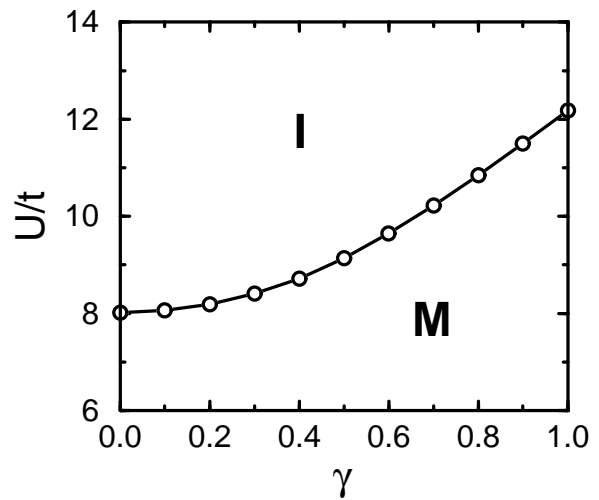}
\caption{
Metal (M) to insulator (I) transition with increasing $U$, as obtained
in the disordered phase as a function of $\gamma$ at $n=1$. At
$\gamma=1$ (orbital case) the transition occurs in the ground state,
while at $\gamma=0$ (spin case) the result of Ref. \onlinecite{Bri70}
is reproduced.\cite{noteafi}
}
\label{fig:mit}
\end{figure}

For the spin model ($\gamma=0$) this problem was solved by Brinkman and
Rice,\cite{Bri70} who showed that an `insulating' state (with $d=0$)
sets in above $U_c\simeq 8t$ (in the present units). It is well
understood by now (see Ref.~\onlinecite{Faz99}) that this mean-field
theory does not give an accurate description of the metal-insulator
transition (in particular it ignores all charge fluctuations in the
insulating phase, where in reality $d\neq 0$).\cite{Fle04} By analogy,
one expects that also in the present case $d\neq 0$ at any finite $U$,
and in fact this follows from the large-$U$ expansion analyzed for the
ordered phases in Secs. \ref{sec:hfco} and \ref{sec:hfre} (for a
disordered phase a similar analysis could also be made). Nevertheless,
the Brinkman-Rice transition from a `metallic' to an `insulating' state
at $n=1$ illustrates nicely the competition between kinetic energy and
Coulomb repulsion energy,\cite{noteafi} and
so it is worthwhile to consider the general case, i.e., with arbitrary
$\gamma$. Then, completely analogously to the spin case, an `insulating'
state is found above $U_c(\gamma)=8|\bar{\epsilon}_0(\gamma)|$.
Similar to what happens upon doping (i.e., at finite $x$) in the
$U=\infty$ limit considered above, here upon allowing double occupancy
(i.e., finite $d$) at $n=1$, the metallic phase gains additional kinetic
energy $\propto\gamma$ due to non-pseudospin-conserving hopping which
lowers the kinetic energy below the value due to pseudospin-conserving
hopping alone (the only one present in the spin case). Therefore, the
metallic phase survives up to a higher value of $U$ than at $\gamma=0$,
as shown in Fig. \ref{fig:mit}.

%%%%%%%%%%%%%%%%%%%%%%%%%%%%%%%%%%%%%%%%%%%%%%%%%%%%%%%%%%%%%%%%%%%%%%%%
%%
%%                         Summary and Conclusions
%%
%%%%%%%%%%%%%%%%%%%%%%%%%%%%%%%%%%%%%%%%%%%%%%%%%%%%%%%%%%%%%%%%%%%%%%%%
\section{ Summary and Conclusions}
\label{sec:summa}

In this paper we have made a detailed analysis of the $e_g$-orbital
Hubbard model on a cubic lattice, exploring the consequences of the
absence of SU(2) symmetry and highlighting them by making a comparison
with the familiar SU(2)-symmetric spin Hubbard model. In the first part
we studied the orbital-ordered phases, of which there is a great
variety, precisely because of the lower symmetry, emphasizing the
difference between the complex-orbital states which retain cubic
symmetry, and the real-orbital states in which cubic symmetry is broken.
Analytical results for the order parameter and the energy of each of
these phases in the HF approximation at large $U/t$ were presented,
demonstrating that the total energy can be conveniently divided into
two contributions: a kinetic energy $\propto t$ given by the $U=\infty$
limit, and a SE contribution $\propto t^2/U$. The SE decides
about the relative stability of the various phases at half-filling,
while the kinetic energy contributes and finally becomes dominant upon
doping. This analytical treatment allowed us:
(i) to demonstrate explicitly that SE contributes in both AO and FO
states,
(ii) to demonstrate that the real-orbital states have their orbitals
aligned with the cubic axes, as well as
(iii) to elucidate the structure of the HF phase diagram for the ordered
phases obtained numerically.\cite{Shi00,Bri01,Mae00}
We emphasize that these properties of $e_g$ orbital degrees of freedom
are essentially different from those of $t_{2g}$ ones, because the
latter satisfy certain symmetries and are thus conserved in the hopping
processes.\cite{Kha00,Har03}

In the second part we investigated the disordered orbital-liquid state.
We have demonstrated that in the strong-correlation limit ($U \gg t$)
indeed orbital (FO or AO) order is not robust for $e_g$ orbitals, and
gets replaced by a disordered (OL) phase, if one goes {\it beyond\/}
the HF approximation and includes electron correlation effects in the
disordered phase as well. This leads us to the conclusion that the HF
results,\cite{Shi00,Mae00,Bri01} suggesting that either the FO+ or the
AO$\pm$ state is realized in a broad range of doping, are particularly
misleading for the orbital Hubbard model. Here the present findings
agree qualitatively with the results of the self-consistent second
order perturbation theory obtained by Kubo and Hirashima.\cite{Kub02}
The situation could be somewhat different in the 2D case, however,
where a tendency towards particular orbital orderings with larger
amplitude of $x^2-y^2$ orbitals is favored by geometry.\cite{Mac99,Tha00}

We considered specifically the $U=\infty$ limit, where the OL competes
with fully polarized ordered phases and we have shown that it is more
stable than any of either uniform FO (\ref{HFFOsc:op}) or staggered AO
(\ref{HFAOsc:op}) states. However, at finite $U$ and for sufficiently
low doping $x$, real-orbital $C$-AO order is stabilized by a
superposition of the SE and the JT effect. Particularly in the regime
of low doping the JT interactions might be stronger than the electronic
interactions of double-exchange type, and the induced orbital order
dictates then the type of magnetic order.\cite{Feh04,Dag04} This regime 
is particularly difficult in realistic models for manganites, as the 
orbital interactions induced by oxygen distortions,\cite{Fei99} and the 
orbital polarization around doped holes\cite{Kil99} give additional 
important contributions and support particular types of orbital order. 
Furthermore, the overall stability of ordered versus disordered (OL) 
phases changes when a realistic Hund's coupling is included.\cite{Mai03} 
It has been shown that the FM phase shrinks then to a range of doping 
$0.2\lesssim x\lesssim 0.5$, the $A$-type AF phase is stable near 
$x=0.5$, while the $C$-AF phase takes over at higher hole doping.

Summarizing, the absence of SU(2) symmetry in the $e_g$-orbital
Hubbard model has severe consequences for the properties of the model
itself and for the stability of orbital-ordered states. The Nagaoka
theorem does not apply to the model of correlated $e_g$ electrons at
$U=\infty$, ordered states are harder to realize than in the spin case,
and the Brinkman-Rice transition occurs at a higher value of $U$. The
qualitatively different properties of the ordered phases show up most
clearly in the {\it inverted stability} (with respect to the spin case)
of the ordered phases with complex orbitals, with ferro (staggered)
orbital order favored at small (large) doping. Most importantly, the
exciting suggestion that such complex-orbital ordered states could be
stable at finite doping\cite{Bri01,Shi00,Mae00} has been disproved,
because of the {\it inherent tendency of $e_g$ systems towards orbital
disorder\/} due to the enhancement of the kinetic energy when SU(2)
symmetry is absent. All these features show that several properties of
spin systems which are usually taken for granted, such as:
(i) the very fact that a ferromagnetic state is an eigenstate of either
an itinerant or the Heisenberg Hamiltonian, and
(ii) the absence of superexchange in ferromagnetic states ---
are in fact the consequences of the SU(2) symmetry of the respective
spin models.

%%%%%%%%%%%%%%%%%%%%%%%%%%%%%%%%%%%%%%%%%%%%%%%%%%%%%%%%%%%%%%%%%%%%%%%%
%%
%%                           ACKNOWLEDGMENTS
%%
%%%%%%%%%%%%%%%%%%%%%%%%%%%%%%%%%%%%%%%%%%%%%%%%%%%%%%%%%%%%%%%%%%%%%%%%
\begin{acknowledgments}
We thank P. Horsch, G. Khaliullin, D. I. Khomskii, J.~Spa\l{}ek,
P. W\"{o}lfle, and particularly K. Ro\'sciszewski for insightful
discussions.
A.~M.~Ole\'s would like to acknowledge support by the Polish State
Committee of Scientific Research (KBN) under Project No.~1 P03B 068 26.
\end{acknowledgments}

%%%%%%%%%%%%%%%%%%%%%%%%%%%%%%%%%%%%%%%%%%%%%%%%%%%%%%%%%%%%%%%%%%%%%%%%
%%
%%                              APPENDIX
%%
%%%%%%%%%%%%%%%%%%%%%%%%%%%%%%%%%%%%%%%%%%%%%%%%%%%%%%%%%%%%%%%%%%%%%%%%
\appendix*

\section{ slave boson representation for real orbitals }

The real-orbital version of the transformation of the electron operators
to slave boson and pseudofermion operators may be derived by making
repeated use of the relation between the real and complex orbitals,
as given by Eqs. (\ref{complex}).
Thus with the real-orbital electron operators given by
\begin{eqnarray}
c_{iz}^{\dagger}&=&\textstyle{\frac{1}{\sqrt{2}}}
        \Big(c_{i+}^{\dagger}+c_{i-}^{\dagger}\Big),      \nonumber \\
c_{ix}^{\dagger}&=&\textstyle{\frac{i}{\sqrt{2}}}
        \Big(c_{i+}^{\dagger}-c_{i-}^{\dagger}\Big),
\label{czx}
\end{eqnarray}
we similarly define the real-orbital pseudofermion operators by
\begin{eqnarray}
f_{iz}^{\dagger}&=&\textstyle{\frac{1}{\sqrt{2}}}
        \Big(f_{i+}^{\dagger}+f_{i-}^{\dagger}\Big),      \nonumber \\
f_{ix}^{\dagger}&=&\textstyle{\frac{i}{\sqrt{2}}}
        \Big(f_{i+}^{\dagger}-f_{i-}^{\dagger}\Big),
\label{fzx}
\end{eqnarray}
while for the slave boson operators we set
\begin{eqnarray}
b_{iz}^{\dagger}&=&\textstyle{\frac{1}{\sqrt{2}}}
        \Big(b_{i+}^{\dagger}+b_{i-}^{\dagger}\Big),    \nonumber \\
b_{ix}^{\dagger}&=&\textstyle{\frac{-i}{\sqrt{2}}}
        \Big(b_{i+}^{\dagger}-b_{i-}^{\dagger}\Big).
\label{bzx}
\end{eqnarray}
Then the fermions (electrons) transform under U(1) rotations as
\begin{eqnarray}
\hat{U}_i^{}(\theta)c_{iz}^{\dagger}\hat{U}_i^{\dagger}(\theta) &=&
  \cos(\theta/2) \: c_{iz}^{\dagger}
- \sin(\theta/2) \: c_{ix}^{\dagger},     \nonumber \\
\hat{U}_i^{}(\theta)c_{ix}^{\dagger}\hat{U}_i^{\dagger}(\theta) &=&
  \sin(\theta/2) \: c_{iz}^{\dagger}
+ \cos(\theta/2) \: c_{ix}^{\dagger},
\label{transczx}
\end{eqnarray}
and similarly for the pseudofermions, while the slave bosons transform
as
\begin{eqnarray}
\hat{U}_i^{}(\theta)b_{iz}^{\dagger}\hat{U}_i^{\dagger}(\theta) =
& &  \cos \theta \: b_{iz}^{\dagger}
+ \sin \theta \: b_{ix}^{\dagger},     \nonumber \\
\hat{U}_i^{}(\theta)b_{ix}^{\dagger}\hat{U}_i^{\dagger}(\theta) =
&-&  \sin \theta \: b_{iz}^{\dagger}
+ \cos \theta \: b_{ix}^{\dagger}.
\label{transbzx}
\end{eqnarray}
The different sign choice in Eq.~(\ref{bzx}) as compared to
Eqs.~(\ref{czx}) and (\ref{fzx}) makes the slave bosons rotate in the
opposite direction as the (pseudo)fermions. This compensates for the
doubled rotation angle in the sense that the transformations are
identical for slave bosons and (pseudo)fermions when $\theta$ is a cubic
angle, and so the pairs
$\{c_{iz}^{\dagger},c_{ix}^{\dagger}\}$,
$\{f_{iz}^{\dagger},f_{ix}^{\dagger}\}$, and
$\{b_{iz}^{\dagger},b_{ix}^{\dagger}\}$ all transform as the $\theta$
and $\epsilon$ component of a cubic $E$ doublet.

Substituting the complex-orbital slave boson representation
(\ref{krbosons}) into Eq.~(\ref{czx}) and applying the inverse
transformations to (\ref{fzx}) and (\ref{bzx}), one obtains the
slave boson representation for the real-orbital fermionic operators
$\{c_{iz}^{\dagger},c_{ix}^{\dagger}\}$ analogous to
Eq.~(\ref{krbosons}). The result is
\begin{eqnarray}
c_{iz}^{\dagger}&=&+\textstyle{\frac{1}{\sqrt{2}}}
                  \Big(b_{iz}^{\dagger}f_{iz}^{\dagger}
                  -b_{ix}^{\dagger}f_{ix}^{\dagger}\Big) e_i ,
   \nonumber \\
c_{ix}^{\dagger}&=&-\textstyle{\frac{1}{\sqrt{2}}}
                  \Big(b_{ix}^{\dagger}f_{iz}^{\dagger}
                  +b_{iz}^{\dagger}f_{ix}^{\dagger}\Big) e_i,
\label{cbfzx}
\end{eqnarray}
corresponding to a representation of the local states by
\begin{eqnarray}
|i0\rangle&=&e_i^{\dagger}|{\rm vac}\rangle,   \nonumber \\
|iz\rangle=c_{iz}^{\dagger}|i0\rangle
     &=& +\textstyle{\frac{1}{\sqrt{2}}}
            \Big(b_{iz}^{\dagger}f_{iz}^{\dagger}
            -b_{ix}^{\dagger}f_{ix}^{\dagger}\Big) |{\rm vac}\rangle,
          \nonumber \\
|ix\rangle=c_{ix}^{\dagger}|i0\rangle
     &=&-\textstyle{\frac{1}{\sqrt{2}}}
            \Big(b_{ix}^{\dagger}f_{iz}^{\dagger}
            +b_{iz}^{\dagger}f_{ix}^{\dagger}\Big) |{\rm vac}\rangle.
\label{krstateszx}
\end{eqnarray}
One recognizes that Eqs.~(\ref{cbfzx}) are indeed the proper expressions
for the $E$ doublet resulting from the product representation
$E \otimes E \otimes A_1$. \cite{Grif}
The expressions (\ref{cbfzx}) are actually even U(1)-invariant, i.e.,
after a rotation in orbital space by an arbitrary angle $\theta$,
they also hold between the fermion operators
$\{c_{iz}^{\prime\dagger},c_{ix}^{\prime\dagger}\}
=\{\hat{U}_i^{}(\theta) c_{iz}^{\dagger} \hat{U}_i^{\dagger}(\theta),
\hat{U}_i^{}(\theta) c_{ix}^{\dagger} \hat{U}_i^{\dagger}(\theta)\}$,
transformed according to Eq.~(\ref{transczx}), and the slave boson and
pseudofermion operators
$\{b_{iz}^{\prime\dagger},b_{ix}^{\prime\dagger}\}$ and
$\{f_{iz}^{\prime\dagger},f_{ix}^{\prime\dagger}\}$,
transformed according to Eqs.~(\ref{transbzx}) and (\ref{transczx}),
respectively. Consequently, since the hopping Hamiltonian (\ref{H_real})
is invariant under a transformation (\ref{transczx}) of the fermion
(electron) operators when $\theta$ is one of the cubic angles
$0,\pm 4\pi/3$ and is accompanied by the corresponding permutation of
the cubic axes, this cubic invariance is retained when the Hamiltonian
is expressed in terms of the slave boson and pseudofermion operators by
means of Eq.~(\ref{cbfzx}).

The constraints given by Eqs.~(\ref{const}) are now replaced by
\begin{eqnarray}
 b_{iz}^{\dagger}b_{iz}^{}+b_{ix}^{\dagger}b_{ix}^{}
    &+&  e_i^{\dagger}e_i^{}=1,
        \nonumber \\
 b_{iz}^{\dagger}b_{iz}^{}+b_{ix}^{\dagger}b_{ix}^{}
&=& f_{iz}^{\dagger}f_{iz}^{}+f_{ix}^{\dagger}f_{ix}^{},
        \nonumber \\
 b_{iz}^{\dagger}b_{ix}^{}-b_{ix}^{\dagger}b_{iz}^{}
&=& f_{iz}^{\dagger}f_{ix}^{}-f_{ix}^{\dagger}f_{iz}^{}.
\label{constzx}
\end{eqnarray}
Again the first constraint excludes double-occupancy, as required in the
limit $U=\infty$, while the last constraint is readily verified to
eliminate the unphysical singly-occupied states,
\begin{eqnarray}
|iA_1\rangle &=&\textstyle{\frac{1}{\sqrt{2}}}
              \Big(b_{iz}^{\dagger}f_{iz}^{\dagger}
              +b_{ix}^{\dagger}f_{ix}^{\dagger}\Big) |{\rm vac}\rangle,
  \nonumber \\
|iA_2\rangle &=&\textstyle{\frac{1}{\sqrt{2}}}
              \Big(b_{ix}^{\dagger}f_{iz}^{\dagger}
              -b_{iz}^{\dagger}f_{ix}^{\dagger}\Big) |{\rm vac}\rangle.
\label{a1a2}
\end{eqnarray}
When the constraints are obeyed rigorously and the unphysical states
strictly projected out, operators connecting the physical and unphysical
subspaces necessarily vanish identically. Specifically one finds
\begin{eqnarray}
b_{iz}^{\dagger}b_{iz}^{}-b_{ix}^{\dagger}b_{ix}^{}=
f_{iz}^{\dagger}f_{iz}^{}-f_{ix}^{\dagger}f_{ix}^{}&=&0,  \nonumber \\
b_{iz}^{\dagger}b_{ix}^{}+b_{ix}^{\dagger}b_{iz}^{}=
f_{iz}^{\dagger}f_{ix}^{}+f_{ix}^{\dagger}f_{iz}^{}&=&0.
\label{zeros}
\end{eqnarray}

It is obvious from the above that the earlier attempt made in Ref.
\onlinecite{Ole00} to construct a real-orbital slave boson
representation by means of
$c_{iz}^{\dagger}=b_{iz}^{\dagger}f_{iz}^{\dagger}e_i$ and
$c_{ix}^{\dagger}=b_{ix}^{\dagger}f_{ix}^{\dagger}e_i$,
followed by renormalization of the slave boson factors by
\begin{eqnarray}
z_{iz}^{\dagger}&=&\frac{b_{iz}^{\dagger}e_i^{}}
{\sqrt{(1-e_i^{\dagger}e_i^{}-b_{ix}^{\dagger}b_{ix}^{})
      (1-b_{iz}^{\dagger}b_{iz}^{})}},
     \nonumber \\
z_{ix}^{\dagger}&=&\frac{b_{ix}^{\dagger}e_i^{}}
{\sqrt{(1-e_i^{\dagger}e_i^{}-b_{iz}^{\dagger}b_{iz}^{})
      (1-b_{ix}^{\dagger}b_{ix}^{})}},
\label{zizx-inc}
\end{eqnarray}
was misguided because it does not conserve the cubic symmetry, and is
thus bound to lead to spurious results. However, also the present
real-orbital representation, though invariant in itself, leaves us
with the problem to construct a proper cubic-invariant renormalization.
This is not straighforward because the hopping Hamiltonian, when
expressed completely in terms of slave boson and pseudofermion operators
referring to `$z$' and `$x$', takes a different appearance for each
cubic axis, like in Eq.~(\ref{H_real}). Moreover, the apparently
plausible renormalization by means of Eqs. (\ref{zizx-inc}) is not
allowed even in combination with the representation (\ref{cbfzx}),
because $z_{iz}^{\dagger}$ and $z_{ix}^{\dagger}$ as defined by
Eqs.~(\ref{zizx-inc}) do not constitute a cubic $E$ doublet as
their denominators are not cubic invariants. Having them replace
$b_{iz}^{\dagger}e_i^{}$ and $b_{ix}^{\dagger}e_i^{}$ in
Eqs.~(\ref{cbfzx}) would spoil also the cubic $E$ doublet nature of the
thus renormalized $c_{iz}^{\dagger}$ and $c_{ix}^{\dagger}$, and so
destroy the cubic symmetry of the Hamiltonian. Equally seriously,
it would also cause the Hamiltonian to commute no longer with the
constraints.

A renormalization not suffering from the above problems and still in
the spirit of the Kotliar-Ruckenstein Ansatz\cite{Kot86} is given by
\begin{eqnarray}
z_{iz}^{\dagger}&=&\frac{b_{iz}^{\dagger}e_i^{}}
{\sqrt{(1-e_i^{\dagger}e_i^{}-\textstyle{\frac{1}{2}} n_{i}^{(b)})
      (1-\textstyle{\frac{1}{2}} n_{i}^{(b)}) }} ,
     \nonumber \\
z_{ix}^{\dagger}&=&\frac{b_{ix}^{\dagger}e_i^{}}
{\sqrt{(1-e_i^{\dagger}e_i^{}-\textstyle{\frac{1}{2}} n_{i}^{(b)})
      (1-\textstyle{\frac{1}{2}} n_{i}^{(b)}) }} ,
\label{zizx}
\end{eqnarray}
where $n_{i}^{(b)}=b_{iz}^{\dagger}b_{iz}^{}+b_{ix}^{\dagger}b_{ix}^{}$.
The mean-field approximation is now made, as in Sec. \ref{sec:krsb},
by replacing only the amplitudes but not the phases by c-numbers.
So, for the offdiagonal two-boson products we set, similarly to
what was done in Eqs. (\ref{bbaroffd}),
\begin{eqnarray}
\langle b_{iz}^{\dagger} e_{i}^{} \rangle &\equiv&
\langle e_{i}^{\dagger} b_{iz}^{} \rangle \equiv
     \bar{b}_{i}\bar{e}_i \: \cos(\alpha_i - \hat{\vartheta}_i) ,
                        \nonumber \\
\langle b_{ix}^{\dagger} e_{i}^{} \rangle &\equiv&
\langle e_{i}^{\dagger} b_{ix}^{} \rangle \equiv
     \bar{b}_{i}\bar{e}_i \: \sin(\alpha_i - \hat{\vartheta}_i) .
\label{bbaroffdreal}
\end{eqnarray}
where $\bar{b}_{i}$ and $\bar{e}_{i}$ are again real quantities.
For the diagonal two-boson products we set
\begin{eqnarray}
\langle b_{iz}^{\dagger} b_{iz}^{} \rangle
  &\equiv&
\langle b_{ix}^{\dagger} b_{ix}^{} \rangle
  \equiv \textstyle{\frac{1}{2}} \bar{b}_{i}^2
                             \nonumber \\
\langle e_{i}^{\dagger} e_{i}^{} \rangle &\equiv& \bar{e}_{i}^2.
\label{bbarrealA1}
\end{eqnarray}
Actually, the real-orbital boson occupation numbers
$n_{iz}^{(b)}=b_{iz}^{\dagger}b_{iz}^{}$ and
$n_{ix}^{(b)}=b_{ix}^{\dagger}b_{ix}^{}$
are not invariants with respect to U(1) rotations, and so one would
prefer to set, in accordance with Eqs. (\ref{bbaroffdreal}), the
corresponding diagonal averages equal to
\begin{eqnarray}
\langle b_{iz}^{\dagger} b_{iz}^{} \rangle
  &\equiv& \bar{b}_{i}^2 \:
  \cos^2 (\alpha_i - \hat{\vartheta}_i) ,
                             \nonumber \\
\langle b_{ix}^{\dagger} b_{ix}^{} \rangle
  &\equiv& \bar{b}_{i}^2 \:
  \sin^2 (\alpha_i - \hat{\vartheta}_i) ,
\label{bbarrealEz}
\end{eqnarray}
in order to make them transform in the same way as the occupation
numbers, by setting also
\begin{eqnarray}
\langle b_{iz}^{\dagger} b_{ix}^{} \rangle \equiv
\langle b_{ix}^{\dagger} b_{iz}^{} \rangle
  \equiv \textstyle{\frac{1}{2}} \bar{b}_{i}^2 \:
   \sin (2 \alpha_i - 2 \hat{\vartheta}_i) .
\label{bbarrealEx}
\end{eqnarray}
However, the expressions (\ref{bbarrealEz}) and (\ref{bbarrealEx})
do not satisfy Eqs. (\ref{zeros}), and so it appears to be impossible to
assign a nontrivial dependence on the phase operator $\hat{\vartheta}_i$
to $\langle b_{iz}^{\dagger} b_{iz}^{} \rangle$ and
$\langle b_{ix}^{\dagger} b_{ix}^{} \rangle$ and yet
simultaneously respect (\ref{zeros}).

The issue is immaterial for carrying out the KR procedure, since
in a state with uniform density, i.e. with $\bar{e}_{i}^2=x$ for all
$i$, it follows from the constraints (\ref{constzx}) that both for
(\ref{bbarrealA1}) and for (\ref{bbarrealEz}) the amplitude satisfies
$\bar{b}_i^2=1-x$, so that
\begin{eqnarray}
\langle z_{iz}^{\dagger} \rangle &\equiv&
\langle z_{iz}^{} \rangle \equiv
   \sqrt{2q(x)} \; \cos ( \alpha_i - \hat{\vartheta}_i )  ,
             \nonumber \\
\langle z_{ix}^{\dagger} \rangle &\equiv&
\langle z_{ix}^{} \rangle \equiv
   \sqrt{2q(x)} \; \sin ( \alpha_i - \hat{\vartheta}_i ) .
\label{zrealav}
\end{eqnarray}
Inserting this into Eqs.~(\ref{cbfzx}) and defining new pseudofermions
by
\begin{eqnarray}
\hat{\hat{f}}_{iz}^{\dagger} &=& \hskip .3cm
  \cos \alpha_i \: \hat{f}_{iz}^{\dagger}
+ \sin \alpha_i \: \hat{f}_{ix}^{\dagger},
         \nonumber \\
                             &=& \hskip .3cm
  \cos (\hat{\vartheta}_i - \alpha_i ) \: f_{iz}^{\dagger}
+ \sin (\hat{\vartheta}_i - \alpha_i ) \: f_{ix}^{\dagger},
         \nonumber  \\
\label{barfrealz}
\hat{\hat{f}}_{ix}^{\dagger} &=&
- \sin \alpha_i \: \hat{f}_{iz}^{\dagger}
+ \cos \alpha_i \: \hat{f}_{ix}^{\dagger},
         \nonumber \\
                             &=& \hskip .3cm
  \sin (\hat{\vartheta}_i - \alpha_i ) \: f_{iz}^{\dagger}
- \cos (\hat{\vartheta}_i - \alpha_i ) \: f_{ix}^{\dagger},
          \nonumber \\
\label{barfrealx}
\end{eqnarray}
where $\{ \hat{f}_{iz}^{\dagger}, \hat{f}_{ix}^{\dagger} \}$ are
related to $\{ \hat{f}_{i+}^{\dagger}, \hat{f}_{i-}^{\dagger} \}$
[see Eq. (\ref{barf})] by Eqs. (\ref{fzx}), one finds that the
mean-field approximation effectively leads to the replacements
\begin{equation}
c_{iz}^{\dagger}\equiv \sqrt{q(x)} \hat{\hat{f}}_{iz}^{\dagger},
\hskip .7cm
c_{ix}^{\dagger}\equiv \sqrt{q(x)} \hat{\hat{f}}_{ix}^{\dagger}.
\label{replace}
\end{equation}
The kinetic part of the Hamiltonian is thus simply renormalized by the
Gutzwiller factor $q(x)$, exactly the same result as obtained in the
complex-orbital approach. As the Hamiltonian is therefore again cubic,
it follows that the resulting real-orbital OL is isotropic
(i.e. $\alpha_i= \pi/4$ at all sites, and
$\langle b_{iz}^{\dagger} b_{iz}^{} \rangle =
 \langle b_{ix}^{\dagger} b_{ix}^{} \rangle = (1-x)/2$),
and identical to the OL obtained in the complex-orbital approach.

%%%%%%%%%%%%%%%%%%%%%%%%%%%%%%%%%%%%%%%%%%%%%%%%%%%%%%%%%%%%%%%%%%%%%%%%
%%
%%                             REFERENCES
%%
%%%%%%%%%%%%%%%%%%%%%%%%%%%%%%%%%%%%%%%%%%%%%%%%%%%%%%%%%%%%%%%%%%%%%%%%


\begin{thebibliography}{00}
\bibitem{notecoh}
        Note that $\theta_i$ is the azimuthal, not the polar angle.
        Here we thus follow the common practice in the orbital
        and Jahn-Teller literature. However, we do not follow the
        Jahn-Teller convention of replacing $-\theta_i/2$ by $\theta_i$,
        which would yield conveniently
        $|i \theta_i\rangle = \cos\theta_i |iz\rangle+
                              \sin\theta_i |ix\rangle$
        but would upset the standard form of the pseudospin vector
        Eq.~(\protect\ref{vector}).
\bibitem{note:JT}
        The complete argument is actually more subtle, because the EQM
        has an offdiagonal matrix element between $|+\rangle$ and
        $|-\rangle$. --- The $x$, $y$, $z$ components of ${\bm T}$ in
        the complex-orbital basis correspond to the pseudospin components
        represented in the conventional real basis by the Pauli matrices
        $\sigma^z$, $\sigma^x$, and $\sigma^y$, respectively, and
        so only $T^x$ and $T^y$ appear in the JT coupling, while $T^z$
        is the electronic part of the conserved pseudo angular momentum;
        compare Ref. \protect \onlinecite{noteJTconv}.
\bibitem{notecon}
        Note that the mixed constraint (\ref{mixconst}) behaves
        differently from the similar constraint satisfied by slave
        bosons alone in Eqs. (\ref{const}), where in mean-field
        approximation the boson amplitudes depend on the actual
        electron density [see Eqs. (\ref{bbarx})]. Here the amplitude
        of the $f_i^{\dagger}$ fermions controls the boson amplitudes
        at finite doping, and the constraint remains observed
        when Eqs. (\ref{bpm}) are used.
\bibitem{noteAOreal}
        Formally the situation at zero doping is anomalous. The equation
        for $\cos\theta_{-}$ remains valid at $x=0$ and so
        $\theta_{-}=\pi/2$, i.e. $\theta_{\rm A}-\theta_{\rm B}=\pi$,
        but $\theta_{+}$ becomes undetermined, implying that the stable
        real AO solution is `antiferro' (${\bm T}^{\rm A}=-{\bm T}^{\rm B}$)
        but that the common axis of the pseudospin vectors has an
        arbitrary direction in the `equatorial' plane. This is due to
        the fact that the angle between the pseudospin vectors is
        determined by the SE energy (see Sections \protect\ref{sec:hfco}
        and \protect\ref{sec:hfre}), while the average direction is
        selected by the kinetic energy, which is zero, though, at $x=0$
        and thus ineffective.
\bibitem{noteU1}
        While $\hat{U}_i(\theta)$ was expressed in
        Eq.~(\protect\ref{utheta}) in terms of the original fermions,
        here it has to be expressed in terms of the slave bosons and
        pseudofermions: $\hat{U}_i(\theta)=\exp( -i\theta T_i^{z(b)} )
                         \centerdot\exp( -i\theta T_i^{z(f)} )$, where
        $T_i^{z(b)}=b_{i+}^{\dagger}b_{i+}^{}-b_{i-}^{\dagger}b_{i-}^{}$
        and  $T_i^{z(f)}=\textstyle\frac{1}{2}
             ( f_{i+}^{\dagger}f_{i+}^{}-f_{i-}^{\dagger}f_{i-}^{} )$.
\bibitem{note-T+T-}
        Therefore the present slave boson representation does not
        simplify the treatment of the Jahn-Teller effect; compare Ref.
        \protect \onlinecite{noteJTconv}.
\bibitem{notepath}
        This prescription is handled more naturally in a path integral
        formulation with the boson field represented in the radial
        gauge, as in Ref.~\protect\onlinecite{Fre92}, and the modulus
        fields replaced by their mean-field value. The redefinition of
        the pseudofermion operators, Eq.~(\protect\ref{barf}), is
        equivalent to the local gauge transformation that eliminates
        the phase fields.
\bibitem{note:slavefermion}
        Equivalent results are therefore obtained for the orbital
        ordered states by a single slave fermion approach.
\bibitem{noteafi}
        Note, however, that due to the Fermi surface instability the AF
        phase, with a gap depending on $U$, has a lower energy in the
        entire range of $U$ (see, e.g., Ref. \protect\onlinecite{Fle04}).
\end{thebibliography}
\end{document}